\documentclass[letter, 12pt, man, natbib, floatsintext, notxfonts, donotrepeattitle]{apa7}\usepackage[]{graphicx}\usepackage[]{xcolor}
\makeatletter
\def\maxwidth{ %
  \ifdim\Gin@nat@width>\linewidth
    \linewidth
  \else
    \Gin@nat@width
  \fi
}
\makeatother

\definecolor{fgcolor}{rgb}{0.345, 0.345, 0.345}
\newcommand{\hlnum}[1]{\textcolor[rgb]{0.686,0.059,0.569}{#1}}%
\newcommand{\hlstr}[1]{\textcolor[rgb]{0.192,0.494,0.8}{#1}}%
\newcommand{\hlopt}[1]{\textcolor[rgb]{0,0,0}{#1}}%
\newcommand{\hlstd}[1]{\textcolor[rgb]{0.345,0.345,0.345}{#1}}%
\newcommand{\hlkwa}[1]{\textcolor[rgb]{0.161,0.373,0.58}{\textbf{#1}}}%
\newcommand{\hlkwb}[1]{\textcolor[rgb]{0.69,0.353,0.396}{#1}}%
\newcommand{\hlkwc}[1]{\textcolor[rgb]{0.333,0.667,0.333}{#1}}%
\newcommand{\hlkwd}[1]{\textcolor[rgb]{0.737,0.353,0.396}{\textbf{#1}}}%

\usepackage{framed}
\makeatletter
\newenvironment{kframe}{%
 \def\at@end@of@kframe{}%
 \ifinner\ifhmode%
  \def\at@end@of@kframe{\end{minipage}}%
  \begin{minipage}{\columnwidth}%
 \fi\fi%
 \def\FrameCommand##1{\hskip\@totalleftmargin \hskip-\fboxsep
 \colorbox{shadecolor}{##1}\hskip-\fboxsep
     \hskip-\linewidth \hskip-\@totalleftmargin \hskip\columnwidth}%
 \MakeFramed {\advance\hsize-\width
   \@totalleftmargin\z@ \linewidth\hsize
   \@setminipage}}%
 {\par\unskip\endMakeFramed%
 \at@end@of@kframe}
\makeatother

\definecolor{shadecolor}{rgb}{.97, .97, .97}
\definecolor{messagecolor}{rgb}{0, 0, 0}
\definecolor{warningcolor}{rgb}{1, 0, 1}
\definecolor{errorcolor}{rgb}{1, 0, 0}
\newenvironment{knitrout}{}{} 

\usepackage{alltt}
\usepackage[english]{babel} 
\usepackage{amsmath, amssymb} 
\usepackage{pdfpages}

\usepackage{times} 
\usepackage[semibold]{sourcesanspro} 
\usepackage{setspace} 
\usepackage[font={small, stretch=1.2}]{caption} 

\usepackage{tikz}
\usetikzlibrary{positioning,shapes,fit}
\definecolor{lightlightgray}{RGB}{211,211,211}
\tikzset{every picture/.style={/utils/exec={\sffamily}}}

\title{Bayesian Approaches to Designing Replication Studies}
\shorttitle{Bayesian Approaches to Designing Replication Studies}

\authorsnames[1,2,1]{Samuel Pawel, Guido Consonni, Leonhard Held}
\authorsaffiliations{{Department of Biostatistics, Center for Reproducible
    Science, University of Zurich}, {Dipartimento di Scienze Statistiche,
    Universit\`{a} Cattolica del Sacro Cuore}}

\leftheader{Pawel, Consonni, Held}

\abstract{Replication studies are essential for assessing the credibility of
  claims from original studies. A critical aspect of designing replication
  studies is determining their sample size; a too small sample size may lead to
  inconclusive studies whereas a too large sample size may waste resources that
  could be allocated better in other studies. Here, we show how Bayesian
  approaches can be used for tackling this problem. The Bayesian framework
  allows researchers to combine the original data and external knowledge in a
  design prior distribution for the underlying parameters. Based on a design
  prior, predictions about the replication data can be made, and the replication
  sample size can be chosen to ensure a sufficiently high probability of
  replication success. Replication success may be defined by Bayesian or
  non-Bayesian criteria, and different criteria may also be combined to meet
  distinct stakeholders and enable conclusive inferences based on multiple
  analysis approaches. We investigate sample size determination in the
  normal-normal hierarchical model where analytical results are available and
  traditional sample size determination is a special case where the uncertainty
  on parameter values is not accounted for. We use data from a multisite
  replication project of social-behavioral experiments to illustrate how
  Bayesian approaches can help design informative and cost-effective replication
  studies. Our methods can be used through the R package BayesRepDesign.}


\authornote{
  \addORCIDlink{Samuel Pawel}{0000-0003-2779-320X}\\
  \addORCIDlink{Guido Consonni}{0000-0002-1252-5926}\\
  \addORCIDlink{Leonhard Held}{0000-0002-8686-5325}

  © 2023, American Psychological Association. This paper is not the copy of
  record and may not exactly replicate the final, authoritative version of the
  article. Please do not copy or cite without authors' permission. The final
  article will be available, upon publication, via its DOI:
  \href{https://doi.org/10.1037/met0000604}{10.1037/met0000604}

  This research has not been preregistered. A preprint has been previously
  published on arXiv \citep{Pawel2022c} and included in the PhD thesis of Samuel
  Pawel \citep{Pawel2023thesis}. We declare that we have no conflicts of
  interest. This work was supported by the Swiss National Science Foundation
  (\#189295). The funder had no role in study design, data collection, data
  analysis, data interpretation, decision to publish, or preparation of the
  manuscript.

  We thank Charlotte Micheloud, Angelika Stefan and Ainesh Sewak for helpful
  comments on drafts of the manuscript. We thank the anonymous reviewer for
  constructive comments. We thank \citet{Protzko2020} for publicly sharing their
  data. Their CC-By 4.0 licensed data were downloaded from
  \href{https://osf.io/42ef9/}{https://osf.io/42ef9/}. The R markdown script
  ``Decline effects main analysis.Rmd'' was executed and the relevant variables
  from the objects ``ES\_experiments'' and ``decline\_effects'' were saved. All
  our analyses were conducted in the R programming language version
  4.3.1 \citep{R}.
  Code to reproduce this manuscript is available at
  \href{https://github.com/SamCH93/BAtDRS}{https://github.com/SamCH93/BAtDRS}
  \citep{Pawel2022github}. A snapshot of the Git repository at the time of
  writing is archived at
  \href{https://doi.org/10.5281/zenodo.7291076}{https://doi.org/10.5281/zenodo.7291076}.
  Methods for Bayesian design of replication studies are implemented in our R
  package BayesRepDesign \citep{Pawel2023pkg} which is available at
  \href{https://CRAN.R-project.org/package=BayesRepDesign}{https://CRAN.R-project.org/package=BayesRepDesign}.

  Correspondence concerning this article should be addressed to Samuel Pawel,
  Epidemiology, Biostatistics and Prevention Institute, University of Zurich,
  Hirschengraben 84, 8001 Zurich, Switzerland. E-mail: samuel.pawel@uzh.ch
}

\DeclareMathOperator{\Nor}{N} 
\newcommand{\given}{\,\vert\,} 
\newcommand{\that}{\hat{\theta}} 
\newcommand{\BFs}{\text{BF}_{\scriptscriptstyle\text{S}}} 
\newcommand{\BFr}{\text{BF}_{\scriptscriptstyle \text{R}}} 
\newcommand{\ps}{p_{\scriptscriptstyle\text{S}}} 

\newcommand{\mutheta}{\mu_{\scriptscriptstyle \theta}}
\newcommand{\sigmatheta}{\sigma_{\scriptscriptstyle \theta}}
\newcommand{\muthatr}{\mu_{\scriptscriptstyle \hat{\theta}_{r}}}
\newcommand{\sigmathatr}{\sigma_{\scriptscriptstyle \hat{\theta}_{r}}}

\newcommand{\zalpha}{z_{\scriptscriptstyle \alpha}}
\newcommand{\zalphatwo}{z_{\scriptscriptstyle \alpha/2}}
\newcommand{\zbeta}{z_{\scriptscriptstyle \beta}}
\newcommand{\bthat}{\hat{\boldsymbol{\theta}}} 
\newcommand{\bsigma}{\boldsymbol{\sigma}} 
\newcommand{\btheta}{\boldsymbol{\theta}} 
\newcommand{\bone}{\mathbf{1}} 

\IfFileExists{upquote.sty}{\usepackage{upquote}}{}
\begin{document}
\maketitle

\section{Introduction}

The replicability of research findings is a cornerstone for the credibility of
science. However, there is growing evidence that the replicability of many
scientific findings is lower than expected \citep{Opensc2015, Camerer2018,
  Errington2021}. This ``replication crisis'' has led to methodological reforms
in various fields of science, one of which is an increased conduct of
replication studies \citep{Munafo2017}. Statistical methodology plays a key role
in the evaluation of replication studies, and various methods have been proposed
for quantifying how ``successful'' a replication study was in replicating the
original finding \citep[among others]{Bayarri2002, Verhagen2014, Simonsohn2015,
  Anderson2016, Patil2016, Johnson2016, Etz2016, vanAert2017, Ly2018, Harms2019,
  Hedges2019, Mathur2020, Held2020, Pawel2020, Bonett2020, Held2021,
  Pawel2022b}. Yet, as with ordinary studies, statistical methodology is not
only important for analyzing replication studies but also for designing them, in
particular for their \emph{sample size determination} (SSD). Optimal SSD is
important since too small sample sizes may lead to inconclusive studies, whereas
too large sample sizes may waste resources which could have been allocated
better in other research projects.

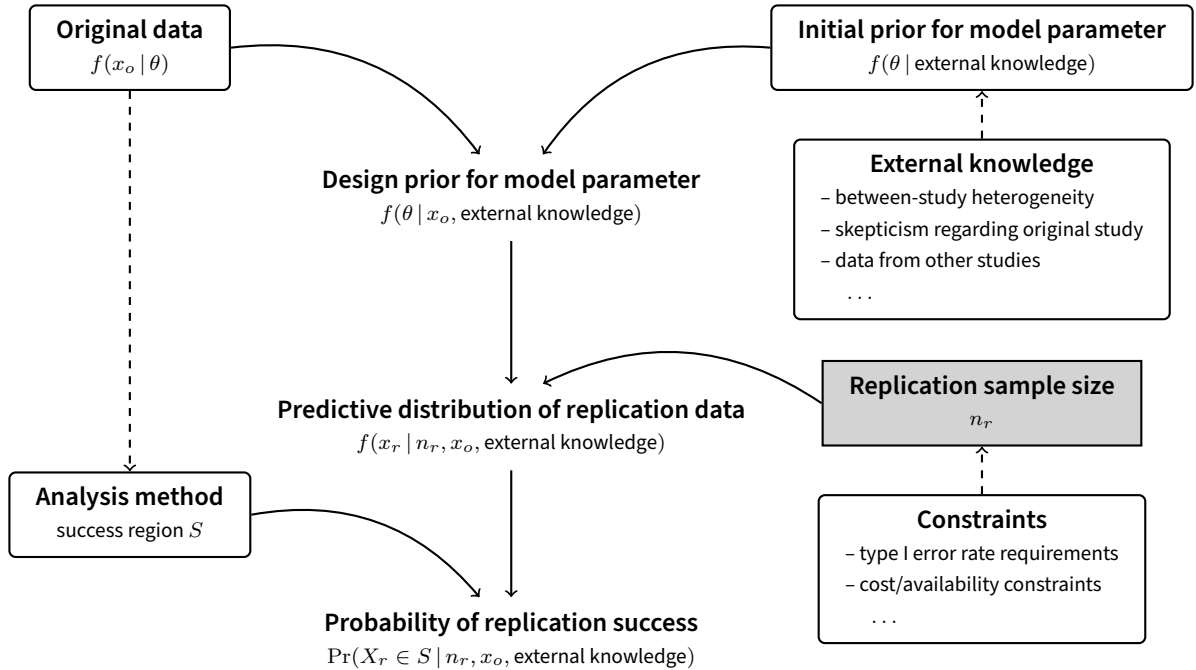
\begin{figure}[!ht]
  \centering

  \begin{tikzpicture}[thick, node distance = 11em and 1em,
  font=\fontsize{10pt}{12pt}\selectfont]

    \node[rectangle, draw, rounded corners = 0.2em] (oData)
    {\begin{tabular}{c} \textbf{Original data} \\
        {\scriptsize $f(x_{o} \given \theta)$} \\
     \end{tabular}};

   \node (designPrior) [below right = 2em and 2em of oData]
   {\begin{tabular}{c} \textbf{Design prior for model parameter} \\
      {\scriptsize $f(\theta \given x_{o}, \text{external
      knowledge})$}\\
    \end{tabular}};

  \node[rectangle, draw, rounded corners = 0.2em] (external) [right = 17em of oData]
  {\begin{tabular}{c} \textbf{Initial prior for model parameter} \\
     {\scriptsize $f(\theta \given \text{external knowledge})$} \\
   \end{tabular}};

   \node[rectangle, draw, rounded corners = 0.2em] (externalknowl) [below = 1.5em of external]
   {\begin{tabular}{l}
      \multicolumn{1}{c}{\textbf{External knowledge}} \\
     {\scriptsize -- between-study heterogeneity} \\
     {\scriptsize -- skepticism regarding original study} \\
     {\scriptsize -- data from other studies} \\
     {\scriptsize $\phantom{\rightarrow} \hdots$} \\
   \end{tabular}};

 \node (rData) [below = 4.5em of designPrior]
 {\begin{tabular}{c} \textbf{Predictive distribution of replication data} \\
    {\scriptsize $f(x_{r} \given n_{r}, x_{o}, \text{external
    knowledge})$}
  \end{tabular}};

\node[rectangle, draw, fill = lightlightgray] (nr) [below = 1.25em of externalknowl]
{\begin{tabular}{c} \textbf{Replication sample size} \\
   {\scriptsize $n_{r}$}
 \end{tabular}};

\node (PRS) [below = 4em of rData]
{\begin{tabular}{c} \textbf{Probability of replication success} \\
   {\scriptsize $\Pr(X_r \in S \given n_{r}, x_{o}, \text{external
    knowledge})$}
 \end{tabular}};

\node[rectangle, draw, rounded corners = 0.2em] (measureRS) [below = 12em of oData]
{\begin{tabular}{c} \textbf{Analysis method}  \\
    {\scriptsize success region $S$} \\
 \end{tabular}};

\node[rectangle, draw, rounded corners = 0.2em] (sideCond) [below = 1.5em of nr]
{\begin{tabular}{l}
   \multicolumn{1}{c}{\textbf{Constraints}} \\
   {\scriptsize -- type I error rate requirements} \\
   {\scriptsize -- cost/availability constraints} \\
   {\scriptsize $\phantom{\rightarrow} \hdots$} \\
 \end{tabular}};

\draw [->] [bend left] (oData.east) to ([xshift=-1em]designPrior.north);
\draw [->] [bend right] (external.west) to ([xshift=1em]designPrior.north);
\draw [->] (designPrior) to (rData);
\draw [->] (rData) to (PRS.north);
\draw [->] [bend right] (nr.west) to ([xshift=1em]rData.north);
\draw [->] [bend left] (measureRS.east) to ([xshift=-1em]PRS.north);
\draw [->] [dashed] (oData) to (measureRS);
\draw [->] [dashed] (sideCond) to (nr.south);
\draw [->] [dashed] (externalknowl.north) to (external.south);

\end{tikzpicture}

  \caption{Schematic illustration of Bayesian sample size determination for
    replication studies. The original and replication data are denoted by
    $x_{o}$ and $x_{r}$, respectively. Both are assumed to come from a
    distribution with density/probability mass function denoted by
    $f(x_{i} \given \theta)$ for $i \in \{o, r\}$. An initial prior with density
    function $f(\theta \given \mathrm{external~knowledge})$ is assigned to the
    model parameter $\theta$.}
\label{fig:SSDschema}
\end{figure}

SSD for replication studies comes with unique opportunities and challenges; the
data from the original study can be used to inform SSD, at the same time the
analysis of replication success based on original and replication study is
typically different from an analysis of a single study for which traditional SSD
methodology was developed. Since the design of replication studies should be
aligned with the planned analysis, a small literature has emerged that
specifically deals with power calculations and SSD for replication studies
\citep{Bayarri2002, Goodman1992, Senn2002, Anderson2017, Micheloud2020,
  vanZwet2022, Held2020, Pawel2022b, Hedges2021, Anderson2022}. However, most of
these articles only deal with selected analysis methods and data models.
An exception is the excellent article by \citet{Anderson2022} which discusses
more general principles of replication SSD in the context of psychological
research, mostly from a frequentist perspective. As they state ``the literature
on Bayesian sample size planning is still nascent, particularly with respect to
Bayes Factors \citep{Schoenbrodt2017}, and has not yet been clearly optimized
for the context of most replication goals'' \citep[p. 18]{Anderson2022}. Our
goal is therefore to complement their article by developing a unified framework
of replication SSD (schematically illustrated in Figure~\ref{fig:SSDschema})
based on principles from Bayesian design approaches \citep{Spiegelhalter1986b,
  Spiegelhalter1986c, Weiss1997, OHagan2001b, Gelfand2002, DeSantis2004,
  Spiegelhalter2004, Schoenbrodt2017, Pek2019, Kunzmann2021, Park2022,
  Grieve2022}. We aim to provide both a theoretical basis for methodologists
developing new methods for design and analysis of replication studies, and also
to illustrate how Bayesian design approaches can practically be used by
researchers planning a replication study.

The design of replication studies is a natural candidate for Bayesian knowledge
updating as it allows to combine uncertain information from different
sources---for instance, the data from the original study and/or expert
knowledge---in a \emph{design prior} distribution for the underlying model
parameters. If the analysis of the replication data is also Bayesian, the design
prior may be different from the \emph{analysis prior} which, unlike the design
prior, is usually desired to be objective or ``uninformative''
\citep{OHagan2001b}. Based on the design prior, predictions about the
replication data can be made and the sample size can be chosen such that the
probability of replication success becomes sufficiently high. Importantly,
Bayesian design approaches can also be used if the planned analysis of the
replication study is non-Bayesian, which is the more common situation in
practice. Bayesian design based on a frequentist analysis is known under various
names, such as ``hybrid classical-Bayesian design'' \citep{Spiegelhalter2004} or
``Bayesian assurance'' \citep{OHagan2005}, and has also been used before for
psychological applications \citep{Pek2019, Park2022} and replication studies
\citep{Anderson2017, Micheloud2020}.

This paper is structured as follows: We start with presenting a general
framework for Bayesian SSD of replication studies which applies to any kind of
data model and analysis method. 
We then investigate design priors and SSD in the normal-normal hierarchical
model framework which provides sufficient flexibility for incorporating the
original data and external knowledge in replication
design. 
No advanced computational methods, such as (Markov Chain) Monte Carlo sampling,
are required for conducting Bayesian SSD in this framework, and in many cases
there are even simple formulae which generalize classical power and sample size
calculations. We illustrate the methodology for several Bayesian and
non-Bayesian analysis methods, and for both singlesite and multisite replication
studies. Since multisite replication studies are becoming increasingly popular
in psychology \citep[e.g.,][]{Klein2018}, we also discuss how to choose the
optimum allocation of samples within and between sites from a Bayesian design
point of view. As a running example we use data from a multisite replication
project of social-behavioral experiments \citep{Protzko2020}. Finally, we close
with concluding remarks, limitations, and open
questions. 

\section{General framework} \label{sec:GFW}

Suppose an original study has been conducted and resulted in a data set $x_{o}$.
These data are assumed to come from a distribution characterized by an unknown
parameter $\theta$ and with density function $f(x_{o} \given \theta)$. To assess
the replicability of a claim from the original study, an independent and
identically designed (apart from the sample size) replication study is conducted
and the goal of the design stage is to determine its sample size $n_{r}$.

As the observed original data $x_{o}$, the yet unobserved replication data
$X_{r}$ are assumed to come from a distribution depending on the parameter
$\theta$. The parameter $\theta$ thus provides a link between the two studies
and the knowledge obtained from the original study can be used to make
predictions about the replication. The central quantity for doing so is the
so-called \emph{design prior} of the parameter $\theta$, which we write as the
posterior distribution of $\theta$ based on the original data and an
\emph{initial prior} for $\theta$
\begin{align}
  \label{eq:dp}
  f(\theta \given x_o, \mathrm{external~knowledge}) =
  \frac{f(x_o \given \theta) \, f(\theta \given \mathrm{external~knowledge})}{f(x_o \given
  \mathrm{external~knowledge})}.
\end{align}
The initial prior of $\theta$ may depend on external knowledge (e.g., data from
other studies) and it represents the uncertainty about $\theta$ before observing
the original data. We will discuss common types of external knowledge in the
replication setting in the next section. 
The design prior~\eqref{eq:dp} hence represents the state of knowledge and
uncertainty about the parameter $\theta$ before the replication is conducted
and, along with an assumed replication sample size $n_{r}$, it can be used to
compute a predictive distribution for the replication data
\begin{align}
  \label{eq:ftr}
  f(x_r \given n_r, x_o, \mathrm{external~knowledge})
  &= \int
    f(x_r \given n_{r}, \theta)
   \, f(\theta \given x_o, \mathrm{external~knowledge})
    \, \mathrm{d}\theta.
\end{align}

After completion of the replication, the observed data $x_r$ will be analyzed in
some way to quantify to what extent the original result could be replicated. The
analysis may involve the original data (e.g., a meta-analysis of the two data
sets) or it may only use the replication data. Typically, there is a
\emph{success region} $S$ which implies that if the replication data are
contained within it ($x_r \in S$), the replication is successful. The
\emph{probability of replication success} can thus be computed by integrating
the predictive density~\eqref{eq:ftr} over $S$. To ensure a sufficiently
conclusive replication design, the sample size $n_{r}$ is determined such that
the probability of replication success is at least as high as a desired target
probability of success, here and henceforth denoted by $1 - \beta$. The required
sample size $n_r^*$ is then the smallest sample size which leads to a
probability of replication success of at least $1 - \beta$, i.e.,
\begin{align}
  \label{eq:nr}
  n_r^* = \inf \left\{n_r : \Pr(X_{r} \in S \given
  n_r, x_{o}, \mathrm{external~knowledge}) \geq 1 - \beta \right\}.
\end{align}

Often, replication studies are analyzed using several methods which quantify
different aspects of replicability and have different success regions (e.g., a
meta-analysis of original and replication data and an analysis of the
replication data in isolation). In this case, the sample size may be chosen such
that the probability of replication success is as high as desired for all
planned analysis methods.

There may sometimes be certain constraints which the replication sample size
needs to satisfy. For instance, in most cases there is an upper limit on the
sample size due to limited resources and/or availability of samples. Moreover,
funders and regulators may also require methods to be \emph{calibrated}
\citep{Grieve2016}, that is, to have appropriate type I error rate control. The
sample size $n_r^*$ may thus also need to satisfy a type I error rate not higher
than some required level.

\section{Sample size determination in the normal-normal hierarchical model}
\label{sec:SS}
We will now illustrate the general methodology from the previous section in the
\emph{normal-normal hierarchical model} where predictive distributions and the
probability of replication success can often be expressed in closed-form,
permitting further insight. 
It is pragmatic to adopt a meta-analytic perspective and use only study level
summary statistics instead of the raw study data since the raw data from the
original study are not always available to the replicators. Typically, the
underlying parameter $\theta$ is a univariate effect size quantifying the effect
on the outcome variable (e.g., a mean difference, a log odds ratio, or a log
hazard ratio). The original and replication study can then be summarized through
an effect estimate $\that$, possibly the maximum likelihood estimate, and a
corresponding standard error $\sigma$, i.e., $x_{o} = \{\that_{o}, \sigma_{o}\}$
and $x_{r} = \{\that_{r}, \sigma_{r}\}$. Effect estimates and standard errors
are routinely reported in research articles or can, under some assumptions, be
computed from $p$-values and confidence intervals. As in the conventional
meta-analytic framework \citep{Sutton2001}, we further assume that for study
$k \in \{o, r\}$ the (suitably transformed) effect estimate $\hat{\theta}_k$ is
approximately normally distributed around a study specific effect size
$\theta_k$ and with (known) variance equal to its squared standard error
$\sigma_k^2$, here and henceforth denoted by
$\hat{\theta}_k \given \theta_k \sim \Nor(\theta_k, \sigma_k^2)$. The standard
error $\sigma_k$ is typically of the form $\sigma_k = \lambda/\surd{n_k}$ with
$\lambda^{2}$ some unit variance and $n_{k}$ the sample size. The ratio of the
original to the replication variance is thus the ratio of the replication to the
original sample size
\begin{align*}
  c = \sigma^2_o/\sigma^2_r = n_r/n_o,
\end{align*}
which is often the main focus of SSD as it quantifies how much the replication
sample $n_{r}$ size needs to be changed compared to the original sample size
$n_{o}$. Depending on the effect size type, this framework might require slight
modifications \citep[see e.g.,][Section 2.4]{Spiegelhalter2004}.

Assuming a normal sampling model for the effect
estimates~\eqref{eq:hat_theta_k}, as described previously, and specifying an
initial hierarchical normal prior for the study specific effect
sizes~\eqref{eq:theta_k} and the effect size~\eqref{eq:theta}, leads to the
normal-normal hierarchical model
\begin{subequations}
\label{eq:hierarch-model}
\begin{align}
  \hat{\theta}_k \given \mspace{-1mu} \theta_k &\sim \Nor(\theta_k, \sigma_k^2)
  \label{eq:hat_theta_k} \\
  \theta_k \given \theta \,\,  &\sim \Nor(\theta, \tau^2) \label{eq:theta_k} \\
  \theta \,\, &\sim \Nor(\mutheta,
  \sigmatheta^2). \label{eq:theta}
\end{align}
\end{subequations}
By marginalizing over the study specific effects sizes, the
model~\eqref{eq:hierarch-model} can alternatively be expressed as
\begin{subequations}
\label{eq:hierarch-model2}
\begin{align}
  \hat{\theta}_k \given \mspace{-1mu} \theta &\sim \Nor(\theta, \sigma_k^2 + \tau^{2})
  \label{eq:hat_theta_k2} \\
  \theta  &\sim \Nor(\mutheta,
  \sigmatheta^2) \label{eq:theta2}
\end{align}
\end{subequations}
which is often more useful for derivations and computations. In the following we
will explain how the normal-normal hierarchical model can be used for SSD of the
replication study.

\subsection{Design prior and predictive distribution}
\label{sec:designpredictive}
The observed original data $x_{o} = \{\that_{o}, \sigma_{o}\}$ can be combined
with the initial prior~\eqref{eq:theta2} 
by standard Bayesian theory for normal prior and likelihood \citep[Section
3.7]{Spiegelhalter2004} to obtain a posterior distribution for the effect size
$\theta$
\begin{align}
  \label{eq:dpnormal}
  \theta \given \hat{\theta}_o, \sigma^2_o
  \sim
  \Nor\left(
  \frac{\hat{\theta}_o}{1 + 1/g} + \frac{\mutheta}{1 + g},
  \frac{\sigma^2_o + \tau^2}{1 + 1/g} \right)
\end{align}
where $g = \sigmatheta^2/(\sigma^2_o + \tau^2)$ is the \emph{relative prior
  variance}. This posterior serves then as the design prior for predicting the
replication data.

It is interesting to contrast the design prior \eqref{eq:dpnormal} to the
``conditional'' design prior \citep{Micheloud2020}, that is, to assume that the
unknown effect size $\theta$ corresponds to the original effect estimate
$\that_{o}$. This is a standard approach in practice, for instance,
\citet{Opensc2015} determined the sample sizes of its 100 replications under
this assumption. In our framework it implies that the normal design
prior~\eqref{eq:dpnormal} becomes a point mass at the original effect estimate
$\that_{o}$, which can either be achieved through overwhelmingly informative
original data ($\sigma^{2}_{o} \downarrow 0$) along with no heterogeneity
($\tau^{2} = 0$), or through an overwhelmingly informative initial prior
($g \downarrow 0$) centered around the original effect estimate
($\mutheta = \that_{o}$). Both cases show that from a Bayesian perspective the
standard approach is unnatural as it either corresponds to making the standard
error $\sigma_{o}$ smaller than it actually was, or to cherry-picking the prior
based on the data.

Based on the design prior~\eqref{eq:dpnormal}, a predictive distribution for the
replication effect estimate $\that_{r}$ can be computed. Specifically, assuming
a replication standard error $\sigma_{r}$ and integrating the marginal density
of the replication effect estimate~\eqref{eq:hat_theta_k2} with respect to the
prior density leads to
\begin{align}
  \label{eq:fthetar}
  \hat{\theta}_r \given \hat{\theta}_o, \sigma^2_o, \sigma^2_r
  \sim
  \Nor\left(\muthatr =
  \frac{\hat{\theta}_o}{1 + 1/g} + \frac{\mutheta}{1 + g}, \sigmathatr^{2} =
  \, \sigma^2_r + \tau^2 + \frac{\sigma^2_o + \tau^2}{1 + 1/g}\right),
\end{align}
which can again be shown using standard Bayesian theory \citep[Section
3.13.3]{Spiegelhalter2004}.
The design prior~\eqref{eq:dpnormal} and the resulting predictive
distribution~\eqref{eq:fthetar}
depend on the parameters of the initial prior ($\tau^ {2}$, $\mutheta$,
$\sigmatheta^{2}$). We will now explain how these parameters can be specified
based on external knowledge.

\subsection{Incorporating external knowledge in the initial prior}
\label{sec:initialPrior}

At least three common types of external knowledge can be distinguished in the
replication setting: (i) expected heterogeneity between original and replication
study due to differences in study design, execution, and population, (ii) prior
knowledge about the effect size either from theory or from related studies,
(iii) skepticism regarding the original study due to the possibility of
exaggerated results.

\subsubsection{Between-study heterogeneity}
\label{sec:heterogeneity}
The expected degree of between-study heterogeneity can be incorporated via the
variance $\tau^2$ in~\eqref{eq:theta_k}. As $\tau^{2}$ decreases, the study
specific effect sizes become more similar, whereas for increasing $\tau^{2}$
they become more unrelated. If the replicators do not expect any heterogeneity
they can thus set $\tau^{2} = 0$ which will lead to the model collapsing to a
common effect model.

If heterogeneity is expected, there are different approaches for specifying
$\tau^{2}$. A domain expert may subjectively assess how much heterogeneity is to
be expected due to the change in laboratory, study population, and other
factors. An alternative is to take an estimate from the literature, e.g., from
multisite replication projects or from systematic reviews. Finally, one can also
specify an upper limit of ``tolerable heterogeneity''. This approach is similar
to specifying a minimal clinically relevant difference in classical power
analysis in the sense that a true replication effect size which is intolerably
heterogeneous from the original effect size is not relevant to be detected. An
absolute \citep[Section 5.7.3]{Spiegelhalter2004} and a relative approach
\citep{Held2020c} can be considered. In the absolute approach, a value of
$\tau^{2}$ is chosen such that a suitable range of study-specific effect sizes
is not larger than an effect size difference considered negligible. For example,
when 95\% of the study specific effect sizes should not vary more than a small
effect size e.g., $d = 0.2$ on standardized mean difference
scale based on the \citet{Cohen1992} effect size classification, this would lead
to $\tau = d/(2 \cdot 1.96) \approx 0.05$. 
In the relative approach, $\tau^{2}$ is specified relative to the variance of
the original estimate $\sigma^{2}_{o}$ using field conventions for tolerable
relative heterogeneity. For example, in the Cochrane guidelines for systematic
reviews \citep{Deeks2019} a value of
$I^{2} = \tau^{2}/(\tau^{2} + \sigma^{2}_{o}) = 40\%$ is classified as
``negligible'', which translates to
$\tau^{2} = \sigma^{2}_{o}/(1/I^{2} - 1) = (2\sigma^{2}_{o})/3$.

We note that one can also assign a prior distribution to $\tau^2$. For an
overview of prior distributions for heterogeneity variances in the normal-normal
hierarchical model see \citet{Rover2021}. In this case there is no closed-form
expression for the predictive distribution of the replication effect estimate
but numerical or Monte Carlo integration need to be used. We illustrate in the
supplement how the probability of replication success can be computed in this
case. The derived closed-form expressions conditional on $\tau^{2}$ are still
useful as they enable computation of the predictive distribution up to a
one-dimensional numerical integration.

\subsubsection{Knowledge about the effect size}
Prior knowledge about the effect size $\theta$ can be incorporated via the prior
mean $\mutheta$ and the prior variance $\sigmatheta^2$ in~\eqref{eq:theta}. For
instance, the parameters may be specified based on a meta-analysis of related
studies \citep{McKinney2021} or based on expert elicitation \citep{OHagan2019}.
The resulting design prior will then contain more information than what was
provided by the original data alone, leading to potentially more efficient
designs. If there is no prior knowledge available, a standard approach is to
specify an (improper) flat prior by letting the variance go to infinity
($\sigmatheta^2 \to \infty$). The resulting design prior will then only contain
the information from the original study.

\subsubsection{Exaggerated original results}
\label{sec:shrinkage}
Potentially exaggerated original results can be counteracted by setting
$\mutheta = 0$ which shrinks the design prior towards smaller effect sizes (in
absolute value) than the observed effect estimate $\that_{o}$. For instance,
replicators could believe that the results from the original study are
exaggerated because there is no preregistered study protocol available. Even
without such beliefs, weakly informative shrinkage priors may also be motivated
from a ``regularization'' point of view as they can correct for statistical
biases \citep{Copas1983, Firth1993} or prevent unreasonable parameter values
from taking over the posterior in settings with uninformative data
\citep{Gelman2009}.

The amount of shrinkage is determined via the prior variance $\sigmatheta^2$. A
flat prior ($\sigmatheta^2 \to \infty$) will lead to no shrinkage, while a
highly concentrated prior ($\sigmatheta^2 \downarrow 0$) will completely shrink
the design prior to a point mass at zero. One option for specifying
$\sigmatheta^2$ is to use an estimate from a corpus of related studies. For
instance, \citet{vanZwet2021} used the Cochrane library of systematic reviews to
specify design priors for hypothetical replication studies of RCTs. If no corpus
is available, a pragmatic alternative is to use the empirical Bayes estimate
based on the original data
\begin{align}
  \label{eq:EBestimate}
  \hat{\sigma}^2_{\scriptscriptstyle{\theta}} = \max\{(\that_{o} - \mutheta)^{2} - \tau^{2} - \sigma^{2}_{o}, 0\}.
\end{align}
The estimate~\eqref{eq:EBestimate} will lead to adaptive shrinkage
\citep{Pawel2020} in the sense that shrinkage is large for unconvincing original
studies (those with small effect estimates in absolute value $|\that_{o}|$
and/or large standard errors $\sigma_{o}$), but disappears as the data become
more convincing (through larger effect estimates in absolute value $|\that_{o}|$
and/or smaller standard errors $\sigma_{o}$).

\subsection{Example: Cross-laboratory replication project}
\label{sec:example}
\begin{figure}[!htb]
\begin{knitrout}\footnotesize
\definecolor{shadecolor}{rgb}{0.969, 0.969, 0.969}\color{fgcolor}
\includegraphics[width=\maxwidth]{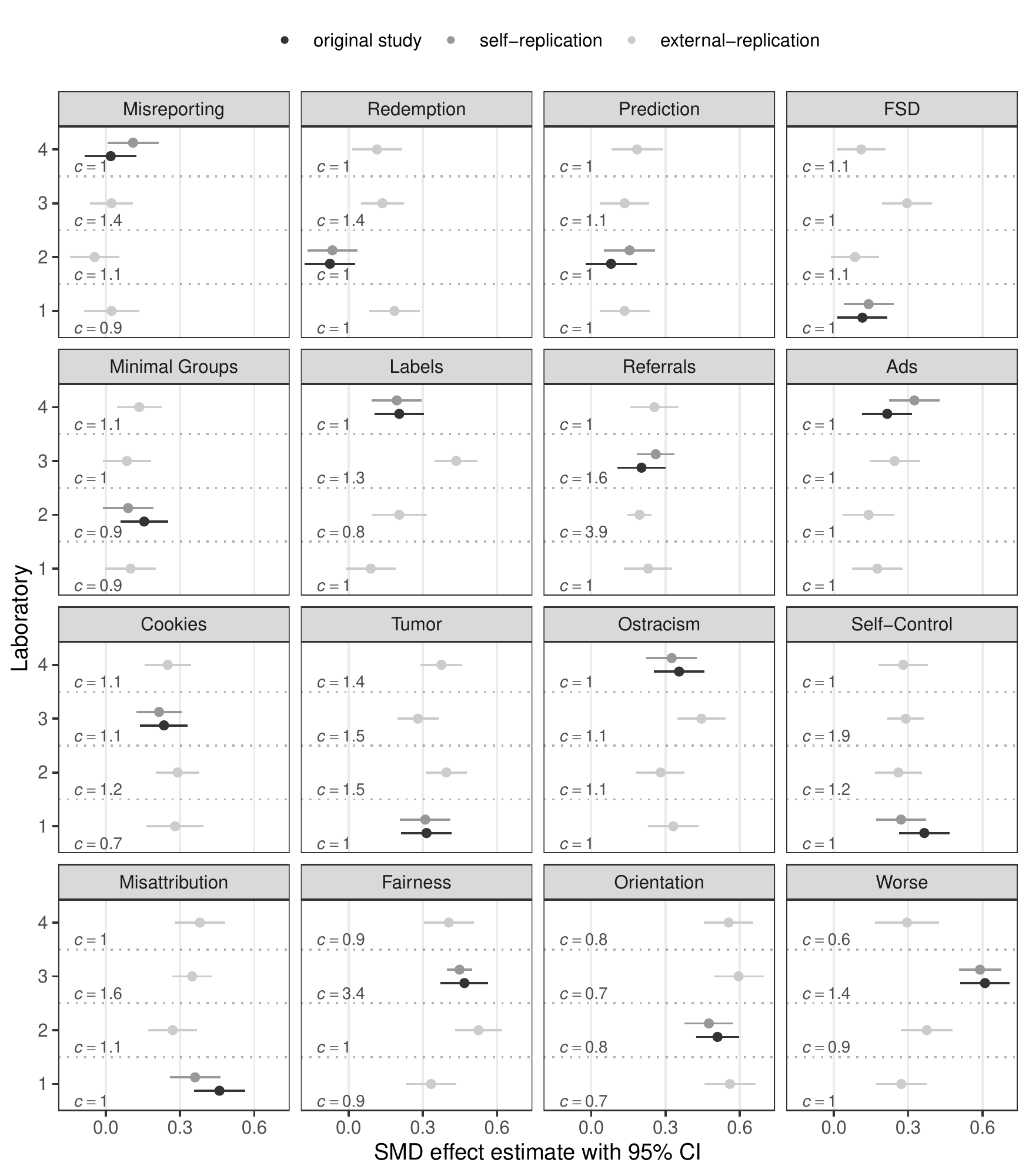} 
\end{knitrout}
\caption{Data from cross-laboratory replication project by \citet{Protzko2020}.
  Shown are standardized mean difference (SMD) effect estimates with 95\%
  confidence intervals stratified by experiment and laboratory. For each
  replication study the relative sample size $c = n_{r}/n_{o}$ 
  is shown. 
}
\label{fig:data}
\end{figure}

We will now illustrate the construction of design priors based on data from a
recently conducted replication project \citep{Protzko2020}, see
Figure~\ref{fig:data} for a summary of the data. The data were collected in four
laboratories over the course of five years and encompassed their typical
social-behavioral experiments on topics such as psychology, communication, or
political science. From the experiments conducted in this period, each lab
submitted four original findings to be replicated. For instance, the original
finding from the ``Labels'' experiment was: ``When a researcher uses a label to
describe people who hold a certain opinion, he or she is interpreted as
disagreeing with those attributes when a negative label is used and agreeing
with those attributes when a positive label is used'' \citep[p.
17]{Protzko2020}, which was based on an effect estimate
$\that_{o} = 0.205$ with 95\% confidence interval from
$0.11$ to $0.3$. For each submitted
original finding, four replication studies were then carried out, one by the
same lab (a \emph{self-replication}) and three by the other three labs (three
\emph{external-replications}).

Most studies used simple between-subject designs with two groups and a
continuous outcome. In this case, the standardized mean difference (SMD) effect
estimate $\that_{i}$ of study $i \in \{o, r\}$ can be computed from the group
means $\bar{y}_{i1}, \bar{y}_{i2}$, group standard deviations $s_{i1}, s_{i2}$,
and group sample sizes $n_{i1}, n_{i2}$ by
\begin{align*}
  \that_{i} = \frac{\bar{y}_{i1} - \bar{y}_{i2}}{s_{i}}
\end{align*}
with
$s^{2}_{i} = \{(n_{i1} - 1)s_{i1}^{2} + (n_{i2} - 1)s_{i2}^{2}\}/(n_{i1} + n_{i2} - 2)$
the pooled sample variance. In the cases where the outcomes were not continuous,
\citet{Protzko2020} transformed the effect estimates to the SMD scale as
explained in their supplementary material. Under a normal sampling model
assuming equal variances in both groups, the approximate variance of $\that_{i}$
is
\begin{align}
  \sigma^{2}_{i} = \frac{n_{i1} + n_{i2}}{n_{i1}n_{i2}} + \frac{\that_{i}^{2}}{2(n_{i1} + n_{i2})}
  \label{eq:varSMD}
\end{align}
\citep{Hedges1981}. A cruder, but more useful approximation for SSD
$\sigma^{2}_{i} \approx 4/n_{i}$ is obtained by assuming the same sample size in
both groups $n_{i1} = n_{i2} = n_{i}/2$, with $n_{i}$ the total sample size, and
neglecting the second term in~\eqref{eq:varSMD} which will be close to zero for
small effect estimates and/or large sample sizes \citep{Hedges2021}. We thus
have the approximate unit variance $\lambda^{2} = 4$ and the relative variance
$c = \sigma^{2}_{o}/\sigma^{2}_{r} = n_{r}/n_{o}$, which can be interpreted as
the ratio of the replication to the original sample size.

Suppose now the original studies have been finished and we want to conduct SSD
for the not yet conducted replication studies. We start by specifying the design
priors (one for each replication). Since the original studies have been
preregistered, we do not expect an exaggeration of their effect estimates due to
selective reporting or other questionable research practices. Therefore, we
choose a flat initial prior for $\theta$ 
which leads to design prior and predictive distribution both centered around the
original effect estimate $\that_{o}$.

For specifying the between-study heterogeneity $\tau$, a distinction needs to be
made between self-replications and external-replications. For self-replications
it is reasonable to set $\tau = 0$ because we would expect no between-study
heterogeneity as the experimental conditions will be nearly identical in both
studies. In contrast, one would expect some between-study heterogeneity for
external-replications as the experimental conditions may slightly differ between
the labs. In the following, we will use $\tau = 0.05$
elicited via the ``absolute'' approach as discussed previously,
so that the range between the 2.5\% and the 97.5\% quantile of the study
specific effect size distribution is equal to a small effect size $d = 0.2$.

\begin{figure}[!ht]
\begin{knitrout}\footnotesize
\definecolor{shadecolor}{rgb}{0.969, 0.969, 0.969}\color{fgcolor}
\includegraphics[width=\maxwidth]{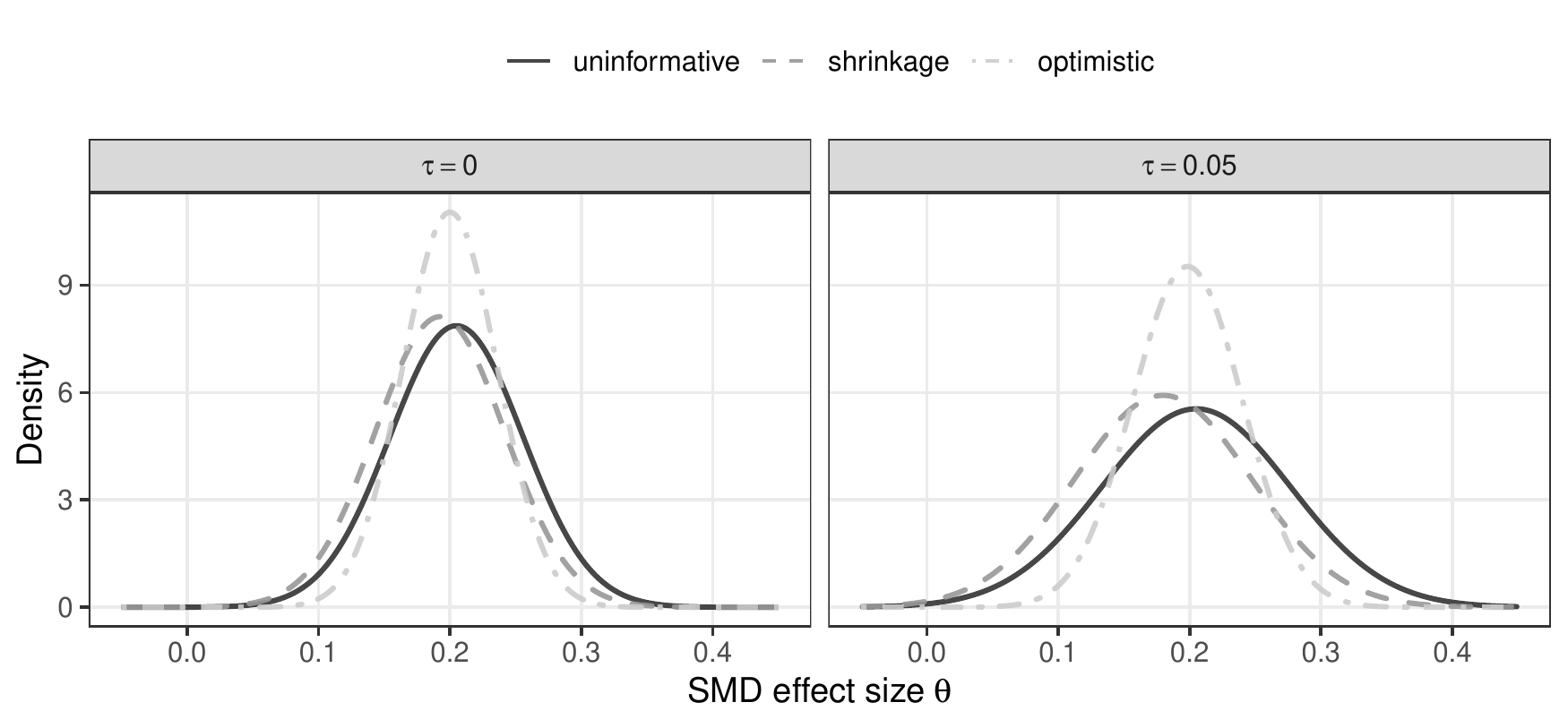} 
\end{knitrout}

\caption{Design priors for the SMD effect size $\theta$ in the
  ``Labels'' experiment based on the original effect estimate
  $\hat{\theta}_{o} = 0.205$ with standard error
  $\sigma_{o} = 0.051$. Shown are different
  choices for the between-study heterogeneity $\tau$ and the initial prior for
  the effect size $\theta$, ``uninformative'' corresponds to a flat prior,
  ``shrinkage'' corresponds to a zero-mean normal prior with empirical Bayes
  variance estimate~\eqref{eq:EBestimate}, and ``optimistic'' corresponds to a
  flat prior updated by the data from a pilot study with effect estimate
  $\hat{\theta}_{p} = 0.195$ and standard error
  $\sigma_{p} = 0.052$.}
\label{fig:dpexample}
\end{figure}

Taken together, we obtain the design prior
$\theta \given \hat{\theta}_{o}, \sigma^{2}_{o} \sim \Nor(\hat{\theta}_{o}, \sigma^{2}_{o})$
for self-replications and the design prior
$\theta \given \hat{\theta}_{o}, \sigma^{2}_{o} \sim \Nor(\hat{\theta}_{o}, \sigma^{2}_{o} + \tau^{2})$
with $\tau^{2} = 0.05^{2}$ for external-replications. For
the ``Labels'' experiment the design prior would be centered around
the original effect estimate $\hat{\theta}_{o} = 0.205$
with variance $\sigma^{2}_{o} = 0.05^{2}$ for a
self-replication, and with variance
$\sigma^{2}_{o} + \tau^{2} = 0.05^{2} + 0.05^{2} \approx 0.07^{2}$
for an external-replication. Figure~\ref{fig:dpexample} (dark-gray solid lines)
shows the densities of the two priors.

While these two priors seem sensible for the \citet{Protzko2020} data, it is
interesting to think about alternative scenarios. If there had been reasons to
believe that the original result might be exaggerated, we could have specified
an initial shrinkage prior. 
For instance, the empirical Bayes estimate for the prior variance
$\sigmatheta^{2}$ from~\eqref{eq:EBestimate} leads to a prior whose mean and
variance are shrunken towards zero by
$12$\% (medium-gray dashed lines in
Figure~\ref{fig:dpexample}). In contrast, if we had prior knowledge about the
effect size $\theta$ from another study, we could have specified an initial
``optimistic'' prior. For example, if the self-replication of the
``Labels'' experiment had been a pilot study and we used its effect
estimate $\hat{\theta}_{p} = 0.195$ and standard
error $\sigma_{p} = 0.05$ to specify the
initial prior, this would lead to a design prior centered around the weighted
mean of original and pilot study, and a prior precision equal to the sum of the
precision of both estimates (light-gray dot-dashed lines in
Figure~\ref{fig:dpexample}). Due to the inclusion of the external data, this
design prior is much more concentrated than the other two.

\subsection{Probability of replication success and required sample size}
To compute the probability of replication success one needs to select an
analysis method and integrate the predictive distribution~\eqref{eq:fthetar}
over the associated success region $S$. There is no universally accepted method
for quantifying replicability and here we do not intend to contribute to the
debate about the most appropriate method. We will simply show the success
regions of different methods and how the replication sample size can be
computed from them. Some methods depend on the direction of the original effect
estimate $\that_{o}$ and throughout we will assume that it was positive
($\that_{o} > 0$). Functions for computing the probability of replication
success and the required sample size are implemented in the R package
BayesRepDesign (see the Appendix) 
for all analysis methods discussed in the following.

\subsubsection{The two-trials rule}
The most common approach for the analysis of replication studies is to declare
replication success when both the original and replication study lead to a
$p$-value for testing the null hypothesis $H_{0} \colon \theta = 0$ smaller than
a pre-specified threshold $\alpha$, usually $\alpha = 5\%$ for two-sided tests
and $\alpha = 2.5\%$ for one-sided tests. This procedure is known as the
\emph{two-trials rule} in drug regulation \citep[Section 12.2.8]{Senn2008}.

We now assume that the one-sided original $p$-value was significant at some
level $\alpha$, i.e., $p_{o} = 1 - \Phi(\that_{o}/\sigma_{o}) \leq \alpha$.
Replication success at level $\alpha$ 
is then achieved if the replication $p$-value is also significant, i.e.,
$p_{r} = 1 - \Phi(\that_{r}/\sigma_{r}) \leq \alpha$, which implies a success
region
\begin{align*}
  S_{\scriptscriptstyle \mathrm{2TR}}
  = \left[\zalpha\, \sigma_{r}, \infty \right),
\end{align*}
where $\zalpha$ is the $1 - \alpha$ quantile of the standard
normal distribution. The probability of replication success is thus given by
\begin{align}
  \label{eq:2TRpros}
  \Pr(\that_{r} \in S_{\scriptscriptstyle \mathrm{2TR}} \given \that_{o}, \sigma_{o}, \sigma_{r})
  =
  \Phi\left(\frac{\muthatr - \zalpha \, \sigma_{r}}{\sigmathatr}\right)
\end{align}
with $\Phi(\cdot)$ the standard normal cumulative distribution function and
$\muthatr$ and $\sigmathatr$ the mean and standard deviation of the predictive
distribution~\eqref{eq:fthetar}. 
Importantly, by decreasing the standard error $\sigma_{r}$ (through increasing
the sample size $n_{r}$), the probability of replication
success~\eqref{eq:2TRpros} cannot become arbitrarily high but is bounded from
above by
\begin{align}
  \label{eq:limP2TR}
  \mathrm{limPr}_{\scriptscriptstyle \mathrm{2TR}} =
  \Phi\left(\frac{ \muthatr}{\sqrt{
\tau^{2} + (\sigma^{2}_{o} + \tau^{2})/(1 + 1/g)}}\right).
\end{align}
The required replication standard error $\sigma_{r}^{*}$ to achieve a target
probability of replication success
$1 - \beta < \mathrm{limPr}_{\scriptscriptstyle \mathrm{2TR}}$ can now be obtained
by equating~\eqref{eq:2TRpros} to $1 - \beta$ and solving for $\sigma_{r}$. This
leads to
\begin{align}
  \label{eq:ssd2tr}
  \sigma_{r}^{*} =
  \frac{\muthatr \zalpha - \zbeta\sqrt{
  (\zalpha^{2} - \zbeta^{2})\left\{\tau^{2} + (\sigma^{2}_{o} +
  \tau^{2})/(1 + 1/g)\right\} + \muthatr^{2}}}{\zalpha^{2} - \zbeta^{2}}
\end{align}
for $\alpha < \beta$. The standard error $\sigma_{r}^{*}$ can subsequently be
translated in a sample size. The translation depends on the type of effect size,
for instance, for SMD effect sizes we can use the approximation
$n_{r}^{*} \approx 4/(\sigma^*_{r})^{2}$ from
earlier. 
Moreover, by assuming a standard error of the form
$\sigma_{r} = \lambda/\surd{n_{r}}$ and plugging in the parameters of the
``conditional'' design prior 
($\tau^{2} = 0$, $\mutheta = \that_{o}$, $g \downarrow 0$), we obtain the
well-known sample size formula \citep[Section 3.3]{Matthews2006}
\begin{align*}
  n_{r}^{*} = \frac{(\zalpha + \zbeta)^{2}}{(\that_{o}/\lambda)^{2}}
\end{align*}
for a one-sided significance test at level $\alpha$ with power $1 - \beta$ to
detect the original effect estimate $\that_{o}$. The formula~\eqref{eq:ssd2tr}
thus generalizes standard sample size calculation to take into account the
uncertainty of the original estimate, between-study heterogeneity and other
types of external knowledge.


\subsubsection{Fixed effect meta-analysis}
The data from the original and replication studies are sometimes pooled via
fixed effect meta-analysis. The pooled effect estimate $\that_{m}$ and standard
error $\sigma_{m}$ are then given by
\begin{align*}
  &\that_{m} =
    \left(\that_{o}/\sigma_{o}^{2} + \that_{r}/\sigma^{2}_{r}\right)\sigma^{2}_{m}&
&\mathrm{and}&                                                                                       &\sigma_{m} = \left(1/\sigma^{2}_{o} + 1/\sigma^{2}_{r}\right)^{-1/2},&
\end{align*}
and they are also equivalent to the mean and standard deviation of a posterior
distribution for the effect size $\theta$ based on the data from both studies
and a flat initial prior for $\theta$. The success region
\begin{align}
  \label{eq:SMA}
  S_{\scriptscriptstyle \mathrm{MA}}
  = \left[\sigma_{r} \zalpha\sqrt{1 + \sigma^{2}_{r}/\sigma^{2}_{o}} -
  (\that_{o} \sigma^{2}_{r})/\sigma^{2}_{o},
  \infty \right)
\end{align}
then corresponds to both replication success defined via a one-sided
meta-analytic $p$-value being smaller than level $\alpha$, i.e.,
$p_{m} = 1 - \Phi(\that_{m}/\sigma_{m}) \leq \alpha$, or to replication success
defined via a Bayesian posterior probability
$\Pr(\theta > 0 \given \that_{o}, \that_{r}, \sigma_{o}, \sigma_{r}) \geq 1 - \alpha$.
Based on the success region~\eqref{eq:SMA} and an assumed standard error
$\sigma_{r}$, the probability of replication success can be computed by
\begin{align}
  \label{eq:MApros}
  \Pr(\that_{r} \in S_{\scriptscriptstyle \mathrm{MA}} \given \that_{o}, \sigma_{o}, \sigma_{r})
  =
  \Phi\left(\frac{\muthatr - \sigma_{r} \zalpha\sqrt{1 + \sigma^{2}_{r}/\sigma^{2}_{o}}
  + (\that_{o} \sigma^{2}_{r})/\sigma^{2}_{o}}{\sigmathatr}\right).
\end{align}
As for the two-trials rule, the probability~\eqref{eq:MApros} cannot be made
arbitrarily high by decreasing the standard error $\sigma_{r}$ but approaches
the limit $\mathrm{limPr}_{\mathrm{2TR}}$ defined in~\eqref{eq:limP2TR}.
The required standard error $\sigma_{r}^{*}$ to achieve a target probability of
replication success $1 - \beta < \mathrm{limPr}_{\mathrm{2TR}}$ can be computed
numerically using root finding algorithms.

\subsubsection{Effect size equivalence test}
\citet{Anderson2016} proposed a method for quantifying replicability based on
effect size equivalence. Under normality, replication success at level $\alpha$
is achieved if the $(1 - \alpha)$ confidence interval for the effect size
difference $\theta_{r} - \theta_{o}$
\begin{align*}
  \that_{r} - \that_{o} \pm \zalphatwo \sqrt{\sigma^{2}_{r} + \sigma^{2}_{o}}
\end{align*}
is fully inside an equivalence region $[-\Delta, \Delta]$ defined via the margin
$\Delta > 0$. This procedure corresponds to rejecting the null hypothesis
$H_{0} \colon |\theta_{r} - \theta_{o}| > \Delta$ in an equivalence test, and it
implies a success region for the replication effect estimate $\that_{r}$ given
by
\begin{align}
  \label{eq:equivalence}
  S_{\scriptscriptstyle \mathrm{E}}
  = \left[\that_{o} - \Delta + \zalphatwo \sqrt{\sigma^{2}_{o} +
  \sigma^{2}_{r}}, \that_{o} + \Delta - \zalphatwo
  \sqrt{\sigma^{2}_{o} + \sigma^{2}_{r}}\right]
\end{align}
for $\Delta \geq \zalphatwo \sqrt{\sigma_{o}^{2} + \sigma^{2}_{r}}$. For too
small margins ($\Delta < \zalphatwo \sqrt{\sigma_{o}^{2} + \sigma^{2}_{r}}$),
the success region~\eqref{eq:equivalence} becomes the empty set meaning that
replication success is impossible. Assuming now that the margin is large enough,
the probability
of replication success can be computed by
\begin{align}
  \label{eq:porsEqu}
  \Pr(\that_{r} \in S_{\scriptscriptstyle \mathrm{E}} \given \that_{o}, \sigma_{o}, \sigma_{r})
  &=
    \Phi\left(\frac{\that_{o} + \Delta - \zalphatwo
  \sqrt{\sigma^{2}_{o} + \sigma^{2}_{r}} -\muthatr}{\sigmathatr}\right) \nonumber \\
&\phantom{=} -
  \Phi\left(\frac{\that_{o} - \Delta + \zalphatwo
  \sqrt{\sigma^{2}_{o} + \sigma^{2}_{r}} -\muthatr}{\sigmathatr}\right).
\end{align}
As with the previous methods, the probability~\eqref{eq:porsEqu} cannot be made
arbitrarily high by decreasing the replication standard error $\sigma_{r}$, but
is bounded by
\begin{align*}
  \mathrm{limPr}_{\scriptscriptstyle \mathrm{E}} =
  \Phi\left(\frac{\that_{o} + \Delta - \zalphatwo\sigma_{o}
  -\muthatr}{\sqrt{\tau^{2} + (\sigma^{2}_{o} + \tau^{2})/(1 + 1/g)}}\right)
  -\Phi\left(\frac{\that_{o} - \Delta + \zalphatwo\sigma_{o}
  -\muthatr}{\sqrt{\tau^{2} + (\sigma^{2}_{o} + \tau^{2})/(1 + 1/g)}}\right).
\end{align*}
The required replication standard error $\sigma_{r}^{*}$ to achieve a target
probability of replication success
$1 - \beta < \mathrm{limPr}_{\scriptscriptstyle \mathrm{E}}$ can again be computed
numerically.

\subsubsection{The replication Bayes factor}
A Bayesian hypothesis testing approach for assessing replication success was
proposed by \citet{Verhagen2014} and further developed by \citet{Ly2018}. They
define a ``replication Bayes factor''
\begin{align*}
  \BFr = \frac{f(x_{r} \given H_{0})}{f(x_{r} \given H_{1})}
\end{align*}
which is the ratio of the marginal likelihood of the replication data $x_{r}$
under the null hypothesis $H_{0} \colon \theta = 0$ to the marginal likelihood
of $x_{r}$ under the alternative hypothesis
$H_{1} \colon \theta \sim f(\theta \given x_{o})$, that is, the posterior of the
effect size $\theta$ based on the original data $x_{o}$. If the original study
provides evidence against the null hypothesis, replication Bayes factor values
$\BFr < 1$ indicate replication success, and the smaller the value the higher
the degree of success.

Under normality and assuming no heterogeneity, the success region for achieving
$\BFr \leq \gamma$ is given by
\begin{align}
  \label{eq:BFrsuccess}
  S_{\scriptscriptstyle \BFr}
  = \left(-\infty, -\sqrt{A} - (\that_{o}\sigma^{2}_{r})/\sigma^{2}_{o}\right] \bigcup
   \left[\sqrt{A} - (\that_{o}\sigma^{2}_{r})/\sigma^{2}_{o}, \infty\right)
\end{align}
with
$A = \sigma^{2}_{r}(1 + \sigma^{2}_{r}/\sigma^{2}_{o}) \{\that_{o}^{2}/\sigma^{2}_{o} - 2 \log \gamma + \log(1 + \sigma^{2}_{o}/\sigma^{2}_{r})\}$.
Details of this calculation are given in the supplement. The fact that the
success region~\eqref{eq:BFrsuccess} is defined on both sides around zero shows
that replication success is also possible if the replication effect estimate
goes in opposite direction of the original one, which is known as the
``replication paradox'' \citep{Ly2018}. The paradox can be avoided using a
modified version of the replication Bayes factor but the success region is no
longer available in closed-form \citep[Appendix D]{Pawel2022b}. Based on the
success region~\eqref{eq:BFrsuccess}, the probability of replication success can
be computed by
\begin{align}
  \label{eq:BFrpros}
  \Pr(\that_{r} \in S_{\scriptscriptstyle \BFr} \given \that_{o}, \sigma_{o}, \sigma_{r})
  &=
      \Phi\left(\frac{\muthatr - \sqrt{A} + (\that_{o}\sigma^{2}_{r})/\sigma^{2}_{o}}{
      \sigmathatr}\right)
    + \Phi\left(\frac{-\sqrt{A} - (\that_{o}\sigma^{2}_{r})/\sigma^{2}_{o}
- \muthatr}{\sigmathatr}\right).
\end{align}
To avoid powering the replication study for the replication paradox, one may
want to compute the probability of replication success only for the part of the
success region with the same sign as the original effect estimate. As for the
other methods, the probability~\eqref{eq:BFrpros} is bounded from above by a
constant
$\mathrm{limPr}_{\scriptscriptstyle \BFr} = \lim_{\sigma_{r} \downarrow 0} \Pr(\that_r \in S_{\scriptscriptstyle \BFr} \given \that_{o}, \sigma_{o}, \sigma_{r})$,
and root finding algorithms can be used to numerically determine the required
standard error $\sigma^{*}_{r}$ for achieving a target probability of
replication success $1 - \beta < \mathrm{limPr}_{\scriptscriptstyle \BFr}$.

\subsubsection{The skeptical \textit{p}-value}
\citet{Held2020} proposed a reverse-Bayes approach for quantifying replication
success. The main idea is to determine the variance of a ``skeptical'' zero-mean
normal prior for the effect size $\theta$ such that its posterior distribution
based on the original study no longer indicates evidence for a genuine effect.
Replication success is then achieved if the replication data are in conflict
with the skeptical prior. The procedure can be summarized by a ``skeptical
$p$-value'' $\ps$, and the lower the $p$-value the higher the degree of
replication success. \citet[Section 2.1]{Held2021} showed that the success region
for replication success defined by $\ps \leq \alpha$ is given by
\begin{align}
  \label{eq:Pssuccess}
  S_{\scriptscriptstyle \ps}
  = \left[\zalpha \sqrt{\sigma^{2}_{r} +
  \frac{\sigma^{2}_{o}}{(z_{o}^{2}/\zalpha^{2}) - 1}}, \infty\right).
\end{align}
From the success region~\eqref{eq:Pssuccess} the probability of replication
success at level $\alpha$ is
\begin{align*}
  \Pr(\that_r \in S_{\scriptscriptstyle \ps} \given \that_{o}, \sigma_{o}, \sigma_{r})
  &=
     \Phi\left(\frac{\muthatr - \zalpha \sqrt{\sigma^{2}_{r} +
    \sigma^{2}_{o}/\{(z_{o}^{2}/\zalpha^{2}) - 1}\}}{\sigmathatr}\right),
\end{align*}
and also bounded from above by a constant
$\mathrm{limPr}_{\scriptscriptstyle \ps} = \lim_{\sigma_{r} \downarrow 0} \Pr(\that_r \in S_{\scriptscriptstyle \ps} \given \that_{o}, \sigma_{o}, \sigma_{r})$.
As for the two-trials rule, the required standard error $\sigma_{r}^{*}$ to
achieve a probability of replication success
$1 - \beta < \mathrm{limPr}_{\scriptscriptstyle \ps}$ can be computed analytically
for $\alpha < \beta$:
\begin{align*}
  \sigma_{r}^{*} = \sqrt{x^{2} - \frac{\sigma^{2}_{o}}{(z_{o}/\zalpha)^{2} - 1}}
\end{align*}
with
\begin{align*}
  x = \frac{\zalpha \muthatr - \zbeta \sqrt{\muthatr^{2} - (\zalpha^{2} - \zbeta^{2})
  [\tau^{2} + (\sigma^{2}_{o} + \tau^{2})/(1 + 1/g) - \sigma^{2}_{o}/\{(z_{o}/\zalpha)^{2}
  - 1\}]}}{\zalpha^{2} - \zbeta^{2}}.
\end{align*}

\subsubsection{The skeptical Bayes factor}
\citet{Pawel2022b} modified the previously described reverse-Bayes assessment of
replication success from \citet{Held2020} to use Bayes factors instead of tail
probabilities as measures of evidence. Again, the procedure can be summarized in
a single quantity termed the ``skeptical Bayes factor'' $\BFs$, with lower
values of $\BFs$ pointing to higher degrees of replication success. The
skeptical Bayes factor is also related to the replication Bayes factor as both
methods use the posterior distribution of $\theta$ based on the data from the
original study as their alternative hypothesis. However, while the replication
Bayes factor uses a point null hypothesis, the skeptical Bayes factor uses a
``skeptical'' zero-mean normal prior for $\theta$ under the null hypothesis
which leads to rather different inferences for replications of unconvincing
original studies \citep[see Section 3 in][]{Pawel2022b}. The success region and
probability of replication success from the skeptical Bayes factor can also be
expressed in closed-form but the derivations are more involved than for the
other methods. For this reason, they are only given in the supplement.

\subsection{Example: Cross-laboratory replication project (continued)}

We will now revisit the ``Labels'' experiment 
and compute the probability of replication success. The parameters of the
analysis methods are specified as follows: For the two-trials rule we use the
conventional one-sided significance level $\alpha = 0.025$,
while for meta-analysis we use the more stringent level
$\alpha = 0.025^{2}$ as the method is based on
two data sets rather than one. We use a $1 - \alpha = 90\%$
confidence interval which is conventionally used in equivalence testing, along
with a margin $\Delta = 0.2$ corresponding to a small SMD effect size
according to the classification from \citet{Cohen1992}. For the skeptical
$p$-value we use the recommended ``golden'' level
$\alpha = 0.062$ as it guarantees that for original studies
which where just significant at $\alpha = 0.025$ replication success is only
possible if the replication effect estimate is larger than the original one
\citep{Held2021}. Finally, for the replication Bayes factor and the skeptical
Bayes factor we use the ``strong evidence'' level
$\gamma = 1/10$ from \citet{Jeffreys1961}.

\begin{figure}[!ht]
\begin{knitrout}\footnotesize
\definecolor{shadecolor}{rgb}{0.969, 0.969, 0.969}\color{fgcolor}
\includegraphics[width=\maxwidth]{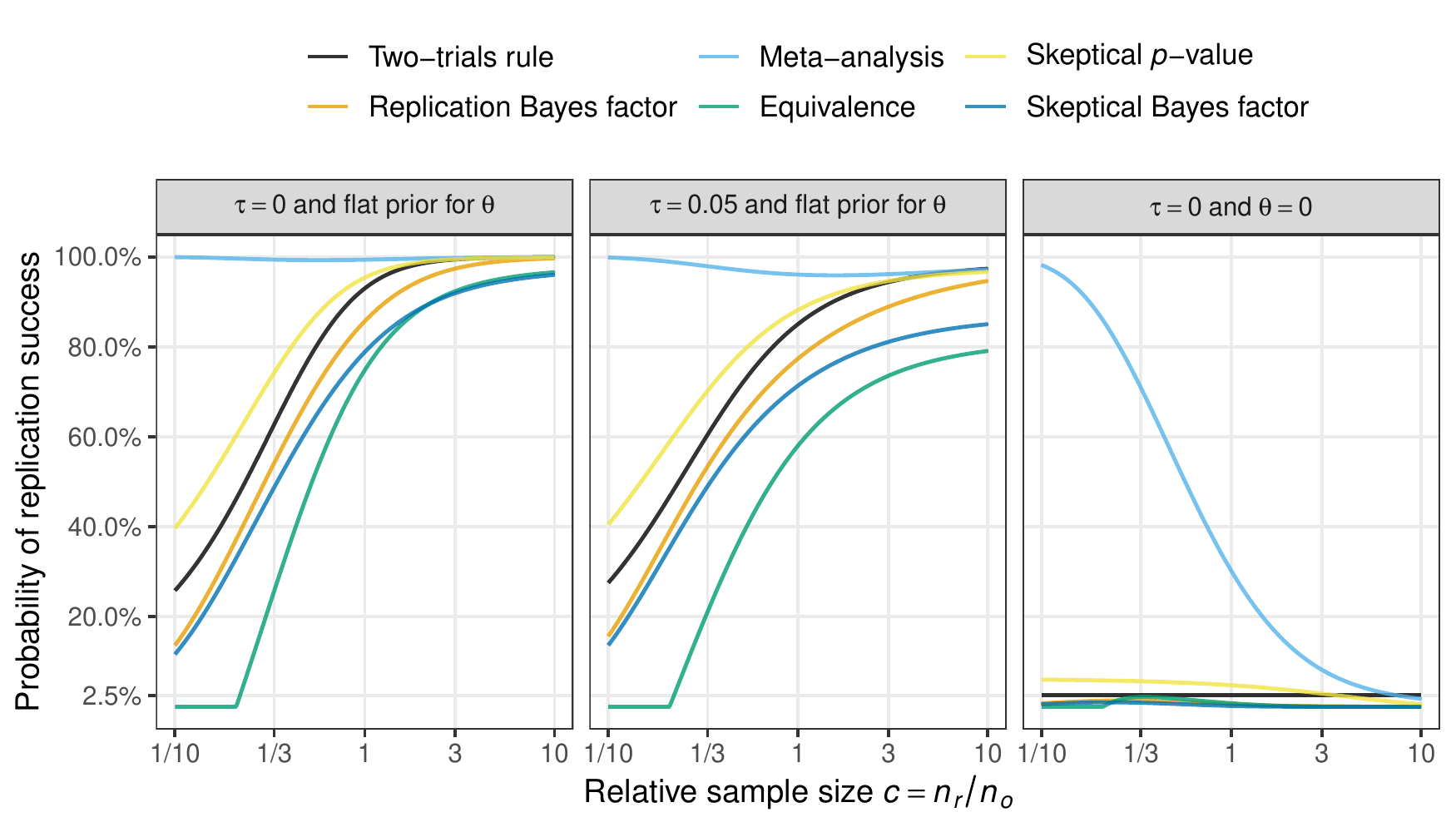} 
\end{knitrout}
\caption{Probability of replication success as a function of the relative sample
  size $c = n_{r}/n_{o}$ for the ``Labels'' experiment with original
  effect estimate $\hat{\theta}_{o} = 0.205$ and
  standard error $\sigma_{o} = 0.051$ under
  different initial prior distributions. 
  Replication success is defined by the two-trials rule at level
  $\alpha = 0.025$, the replication Bayes factor at level
  $\gamma = 1/10$, fixed effect meta-analysis at level
  $\alpha = 0.025^{2}$, effect size equivalence based on
  $90\%$ confidence interval and with margin
  $\Delta = 0.2$, skeptical $p$-value at level
  $\alpha = 0.062$, and skeptical Bayes factor at level
  $\gamma = 1/10$.}
\label{fig:example}
\end{figure}

Figure~\ref{fig:example} shows the probability of replication success as a
function of the relative sample size $c = n_{r}/n_{o}$ and for different initial
priors. The left and middle plot are based on a flat initial prior for the
effect size 
without heterogeneity ($\tau = 0$) and with heterogeneity
($\tau = 0.05$), respectively. The right plot shows the
prior corresponding to the ``fixed effect null hypothesis''
$H_{0} \colon \theta = 0 ~\text{and}~ \tau^{2} = 0$, 
so that the probability of replication success is the type I error rate which
some stakeholders might require to be ``controlled'' at some adequate level.

We see from the left and middle plots that increasing the relative sample size
monotonically increases the probability of replication success for all methods
but meta-analysis (light blue). Meta-analysis shows a non-monotone behavior
because the original study was already highly significant so that the pooled
effect estimate is significant even for replication studies with very small
sample size \citep{Micheloud2020}. The uncertainty regarding the replication
effect estimate $\that_{r}$ may therefore even reduce the probability of
replication success for meta-analysis if the sample size is increased. If
heterogeneity is taken into account (middle plot) the probability of replication
success becomes closer to 50\% for all methods except the equivalence test,
reflecting the larger uncertainty about the effect size $\theta$. To achieve
$80\%$ probability of replication success the fewest
samples are required with meta-analysis, followed by the skeptical $p$-value,
the two-trials rule, the replication Bayes factor, the skeptical Bayes factor,
and lastly the equivalence test. If the sample size should guarantee a
sufficiently conclusive replication study with all these methods, the
replication sample size has to be slightly larger than the original one if no
heterogeneity is assumed ($\tau = 0$), while it has to be increased more than
ten-fold if heterogeneity is assumed ($\tau = 0.05$).
However, this is mostly due to the equivalence test which requires by far the
most samples. If the equivalence test sample size is ignored, the relative
sample size $c = 2.5$ ensures at least
$80\%$ probability of replication success under
heterogeneity with the remaining methods.

The right plot in Figure~\ref{fig:example} shows that the type I error rate of
the two-trials rule (black) stays constant at
$\alpha = 0.025$, as expected by definition of the method.
In contrast, the type I error rates of the other methods vary with the relative
sample size $c$ but most of them stay below
$\alpha = 0.025$ for all $c$ with the exception of
meta-analysis and the skeptical $p$-value. Meta-analysis (light blue) has an
extremely high type I error rate as the pooling with the highly significant
original data leads to replication success if the replication sample size is not
drastically increased. The type I error rate of the skeptical $p$-value (yellow)
is only slightly higher than $\alpha = 0.025$ which is
expected since the level $\alpha = 0.062$ is used for
declaring replication success with the skeptical $p$-value, and its type I error
rate is always smaller than the level for thresholding it \citep{Held2020}. The
type I error rate of the skeptical $p$-value decreases to values smaller than
$\alpha = 0.025$ of the two-trials rule at approximately
$c = 3$.

\begin{figure}[!htb]
\begin{knitrout}\footnotesize
\definecolor{shadecolor}{rgb}{0.969, 0.969, 0.969}\color{fgcolor}
\includegraphics[width=\maxwidth]{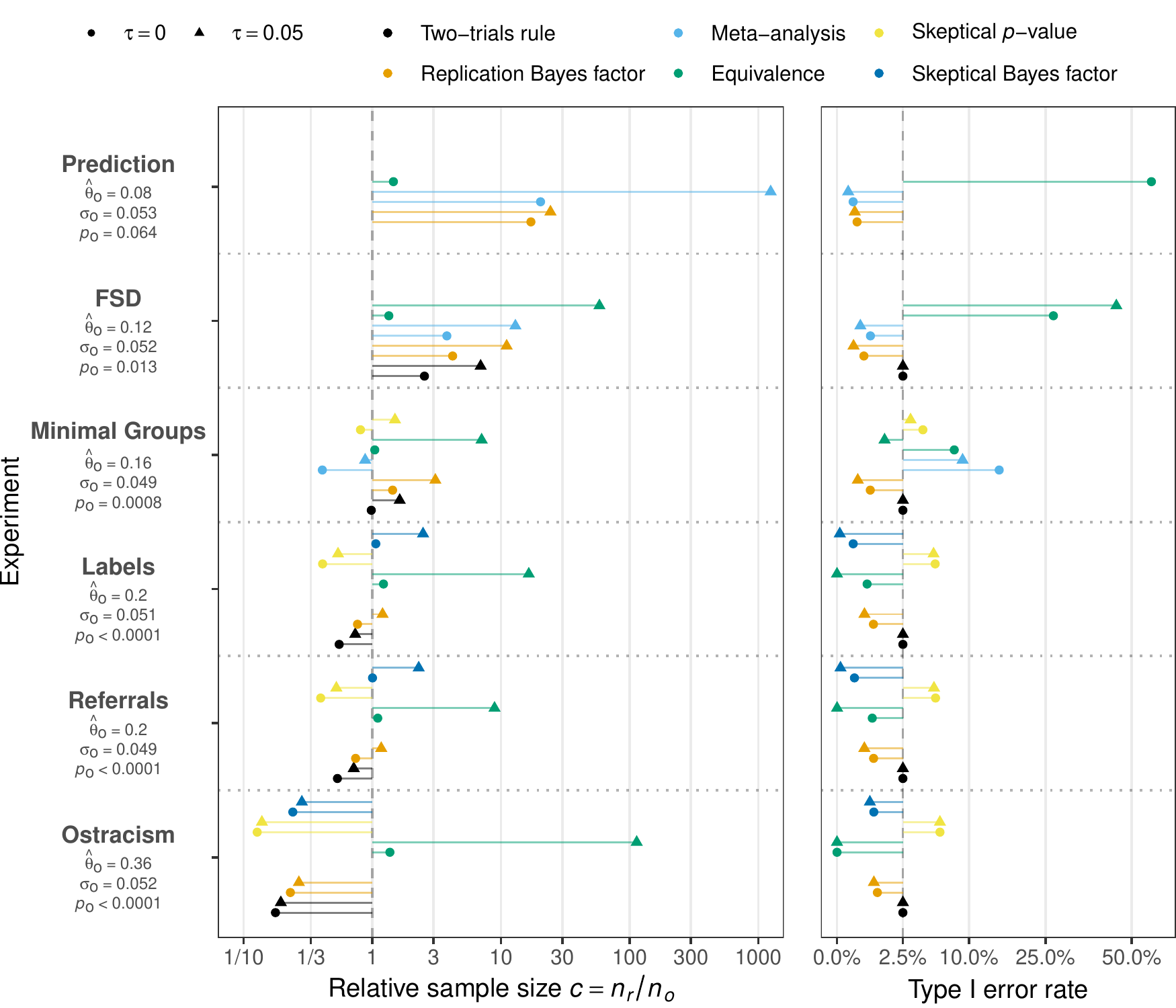} 
\end{knitrout}
\caption{The left plot shows the required relative sample size $c = n_r/n_o$ to
  achieve a target probability of replication success of
  $1 - \beta = 80$\% (if possible). Replication success
  is defined through the two-trials rule at level $\alpha = 0.025$,
  replication Bayes factor at level $\gamma = 1/10$, fixed
  effect meta-analysis at level $\alpha = 0.025^{2}$,
  effect size equivalence at level $\alpha = 0.1$ with margin
  $\Delta = 0.2$, skeptical $p$-value at level
  $\alpha = 0.062$, and skeptical Bayes factor at level
  $\gamma = 1/10$ for an illustrative subset of studies from the
  \citet{Protzko2020} replication project, the supplement shows results for all
  studies. A flat initial prior 
  is used for the effect size $\theta$ either without ($\tau = 0$) or with
  heterogeneity ($\tau = 0.05$). The right plot shows the
  type I error rate associated with the required sample size. Experiments are
  ordered (top to bottom) by their original one-sided $p$-value.
}
\label{fig:ssd-all}
\end{figure}

We now perform SSD for an illustrative subset of studies from the
\citet{Protzko2020} replication project. Figure~\ref{fig:ssd-all} shows the
required relative sample size and the associated type I error rates if a sample
size can be computed for a target probability of replication success of
$1 - \beta = 80$\%. If there is no sample size for
which a probability of $80$\% can be achieved, the
space is left blank. For example, in the application of the meta-analysis method
to the ``Labels'' experiment the probability remains above
$80$\% for any relative sample size and therefore no
sample size is shown.

We see that for all methods except the equivalence test, the required relative
sample size $c$ decreases as the original $p$-value $p_{o}$ decreases and
original studies with very small $p$-values require much fewer samples in the
replication study. For example, in the ``Ostracism'' experiment with
$p_{o} < 0.0001$ the required sample size for all methods except the equivalence
test is at most one-third the size of the original. For the equivalence test the
required sample size depends instead on the size of the original standard error
$\sigma_{o}$ and smaller standard errors lead to smaller required sample sizes
in the replication. For example, the ``Referrals'' experiment with original
standard error $\sigma_{o} = 0.049$ requires fewer samples for the equivalence
test than the ``Ostracism'' experiment with original standard error
$\sigma_{o} = 0.052$.

Figure~\ref{fig:ssd-all} also shows that accounting for heterogeneity
(triangles) increases the required sample size for all methods compared to
ignoring it (points). Although more costly to the researcher, larger sample
sizes also reduce the type I error rate for most methods (right plot). Comparing
the type I error rates of the different methods, we see again the pattern that
the type I error rates of the equivalence test and the skeptical $p$-value are
higher than the type I error rate of 2.5\% of the two-trials rule. However,
while the type I error rate of the skeptical $p$-value decreases when
replication studies require larger samples sizes, the type I error rate of the
equivalence test may also be high if the replication requires very large sample
sizes (e.g., for the ``Fast Social Desirability (FSD)'' experiment), since it depends on whether the
original effect estimate $\that_{o}$ is sufficiently different from zero. If the
original effect estimate $\that_{o}$ is close to zero, the type I error rate of
the equivalence test increases drastically, since equivalence can be established
even if the original and replication effect estimates are close to zero.

The supplement shows the same analysis for all studies in the
\citet{Protzko2020} project. Most original studies were highly significant and
therefore require fewer samples in the replication than in the original study to
achieve a target probability of $1 - \beta = 80$\% for
replication success with all methods except the equivalence test. Some original
studies were less convincing and therefore require larger replication sample
sizes. The additional samples needed for these studies could be reallocated from
the studies that require fewer samples. The project would still use the same
total sample size, but it would be more efficiently allocated. An exception to
this conclusion is the equivalence test which in most cases requires larger
replication sample sizes. This is because the original standard errors of all
studies are relatively large compared to the specified equivalence margin.
Therefore, if one plans to analyze the original and replication pair with an
equivalence test, this should already be taken into account at the design stage
of the original study, since an imprecise original study will diminish the
chances of replication success with this method.

\subsection{Sample size determination for multisite replication projects}
So far we considered the situation where a pair of a single original and a
single replication study are analyzed in isolation. However, if multiple
replications per single original study are conducted (\emph{multisite}
replication studies), the ensemble of replications can also be analyzed jointly.
In this case, some adaptations of the SSD methodology are required.

The replication effect estimate and its standard error are now vectors
$\bthat_{r} = (\that_{r1}, \dots, \that_{rm})^{\top}$ and
$\bsigma^{2}_{r} = (\sigma^{2}_{r1}, \dots, \sigma^{2}_{rm})^{\top}$ consisting
of $m$ replication effect estimates and their standard errors. The normal
hierarchical model for the replication estimates $\bthat_{r}$ then becomes
\begin{subequations}
\label{eq:hierarch-model2}
\begin{align}
  \bthat_r \given \mspace{-1mu} \btheta_r &\sim \Nor_{m}\left\{\btheta_r, \mathrm{diag}\left(\bsigma_r^2\right)\right\}
  \label{eq:hat_theta_k22} \\
  \btheta_r \given \theta \,\,  &\sim \Nor_{m}\left\{\theta \, \bone_{m},
  \tau^2 \mathrm{diag}(\bone_{m})\right\}, \label{eq:theta_k2}
\end{align}
\end{subequations}
where $\btheta_{r}$ is a vector of $m$ study specific effect sizes, $\bone_{m}$
is a vector of $m$ ones, and $\Nor_{m}(\boldsymbol{\mu}, \boldsymbol{\Sigma})$
denotes the $m$-variate normal distribution with mean vector $\boldsymbol{\mu}$
and covariance matrix $\boldsymbol{\Sigma}$. By marginalizing over the study
specific effect size $\btheta_{k}$, the model can alternatively be expressed by
\begin{align}
  \bthat_r \given \theta \sim \Nor_{m}\left\{\theta \, \bone_{m},
\mathrm{diag}\left(\bsigma_r^2 + \tau^2 \bone_{m}\right)\right\},
  \label{eq:margimulti}
\end{align}
so the predictive distribution of $\bthat_r$ based on the design
prior~\eqref{eq:dpnormal} is given by
\begin{align}
  \bthat_r \given \hat{\theta}_o, \sigma^{2}_{o}, \bsigma^{2}_{r}
  \sim \Nor_{m}\left\{\muthatr \bone_{m},
  \mathrm{diag}\left(\bsigma_r^2 + \tau^2 \bone_{m}\right) +
  \left(\frac{\tau^{2} + \sigma^{2}_{o}}{1 + 1/g}\right)
  \bone_{m} \bone_{m}^{\top} \right\}
  \label{eq:ftrmulti}
\end{align}
with $ \muthatr$ the mean of the predictive distribution of a single replication
effect estimate from~\eqref{eq:fthetar}. Importantly, the replication effect
estimates are correlated as the covariance matrix in~\eqref{eq:ftrmulti} has
$(\tau^2+\sigma^{2}_{o})/(1 + 1/g)$ in the off-diagonal entries.

Often the assessment of replication success can be formulated in terms of a
weighted average of the replication effect estimates
$\hat{\theta}_{r*} = (\sum_{i = 1}^{m} w_{i} \hat{\theta}_{ri})/(\sum_{i = 1}^{m} w_{i})$
with $w_{i}$ the weight of replication $i$. For instance, several multisite
replication projects \citep[e.g.,][]{Klein2018} have defined replication success
by the fixed or random effect(s) meta-analytic effect estimate of the replication
effect estimates achieving statistical significance. Based on the predictive
distribution of the replication effect estimate vector~\eqref{eq:ftrmulti}, the
predictive distribution of the weighted average $\hat{\theta}_{r*}$ is given by
\begin{align}
  \hat{\theta}_{r*} \given \hat{\theta}_o, \sigma^{2}_{o}, \bsigma^{2}_{r}
  \sim \Nor\Biggl\{
  \mu_{\scriptscriptstyle \that_{r}},
  \sigma^{2}_{\scriptscriptstyle \that_{r*}} =
  \biggl(\sum_{i=1}^{m} w_{i}^{2} \sigma^{2}_{\scriptscriptstyle \that_{ri}} +
  \sum_{i=1}^{m}\sum_{\substack{j =1 \\ j\neq i}}^{m} w_{i}w_{j}
\frac{\tau^{2} + \sigma^{2}_{o}}{1 + 1/g} \biggr)\big/\biggl(\sum_{i=1}^{m} w_{i}\biggr)^{2} \Biggr\}
  \label{eq:ftwm}
\end{align}
with $\sigma^{2}_{\scriptscriptstyle \that_{ri}}$ the predictive variance of a
single replication effect estimate with standard error $\sigma_{ri}$ as
in~\eqref{eq:fthetar}. In particular, when the studies receive equal weights
($w_{i} = w$ for $i = 1, \dots, m$) and the standard errors of the replication
effect estimates are equal ($\sigma_{ri} = \sigma_{r}$ for $i = 1, \dots, m$),
the predictive variance becomes
\begin{align}
  \sigma^{2}_{\scriptscriptstyle \that_{r*}} =
    \frac{\sigma^{2}_{r} + \tau^{2}}{m} + \frac{\tau^{2} + \sigma^{2}_{o}}{1 + 1/g}.
  \label{eq:predvarequal}
\end{align}
The probability of replication success can now be obtained by
integrating~\eqref{eq:ftrmulti} or \eqref{eq:ftwm} over the corresponding
success region $S$. This may be more involved if the success region is defined
in terms of the replication effect estimate vector $\bthat_{r}$, whereas it is
as simple as in the singlesite replication case if the success region is
formulated in terms of the weighted average $\that_{r*}$.

\subsubsection{Optimal allocation within and between sites}
A key challenge in SSD for multisite replication studies is the optimal
allocation of samples within and between sites, that is, how many sites $m$ and
how many samples $n_{ri}$ per site $i$ should be used. A similar problem exists
in SSD for cluster randomized trials and we can adapt the common solution based
on cost functions \citep{Raudenbush1997}. 
The optimal configuration is determined so that the probability of replication
success is maximized subject to a constrained cost function which accounts for
the (typically different) costs of additional samples and sites.

For example, assume a balanced design ($n_{ri} = n_{r}$ for $i = 1, \dots, m$)
and that the standard errors of the replication effect estimates are inversely
proportional to the square-root of the sample size
$\sigma_{ri} = \lambda/\surd{n_{r}}$ for some unit variance $\lambda^{2}$.
Further, assume that maximizing the probability of replication success
corresponds to minimizing the variance of the weighted average
$\sigma^{2}_{\scriptscriptstyle \that_{r*}}$ in~\eqref{eq:predvarequal}. Let
$K_{s}$ denote the cost of an additional site, and $K_{c}$ the cost of an
additional sample/case. The total cost of the project is then
$K = m(K_{c} \, n_{r} + K_{s})$, and constrained minimization of the predictive
variance~\eqref{eq:predvarequal} leads to the optimal sample size per site
\begin{align*}
  n_{r}^{*} =  \frac{\lambda}{\tau} \, \sqrt{\frac{K_{s}}{K_{c}}}
\end{align*}
which is equivalent to the optimal cluster sample size known from cluster
randomized trials \citep{Raudenbush2000}. Note that the optimal sample size per
site may be different for other analysis approaches where maximizing the
probability of replication success does not correspond to minimizing the
variance of the weighted average. Moreover, there are also practical
considerations which affect the choice of how many sites should be included in a
project. For instance, there may simply not be enough labs available with the
required expertise to perform the replication experiments.

\subsection{Example: Cross-laboratory replication project (continued)}
\begin{figure}[!htb]
\begin{knitrout}\footnotesize
\definecolor{shadecolor}{rgb}{0.969, 0.969, 0.969}\color{fgcolor}
\includegraphics[width=\maxwidth]{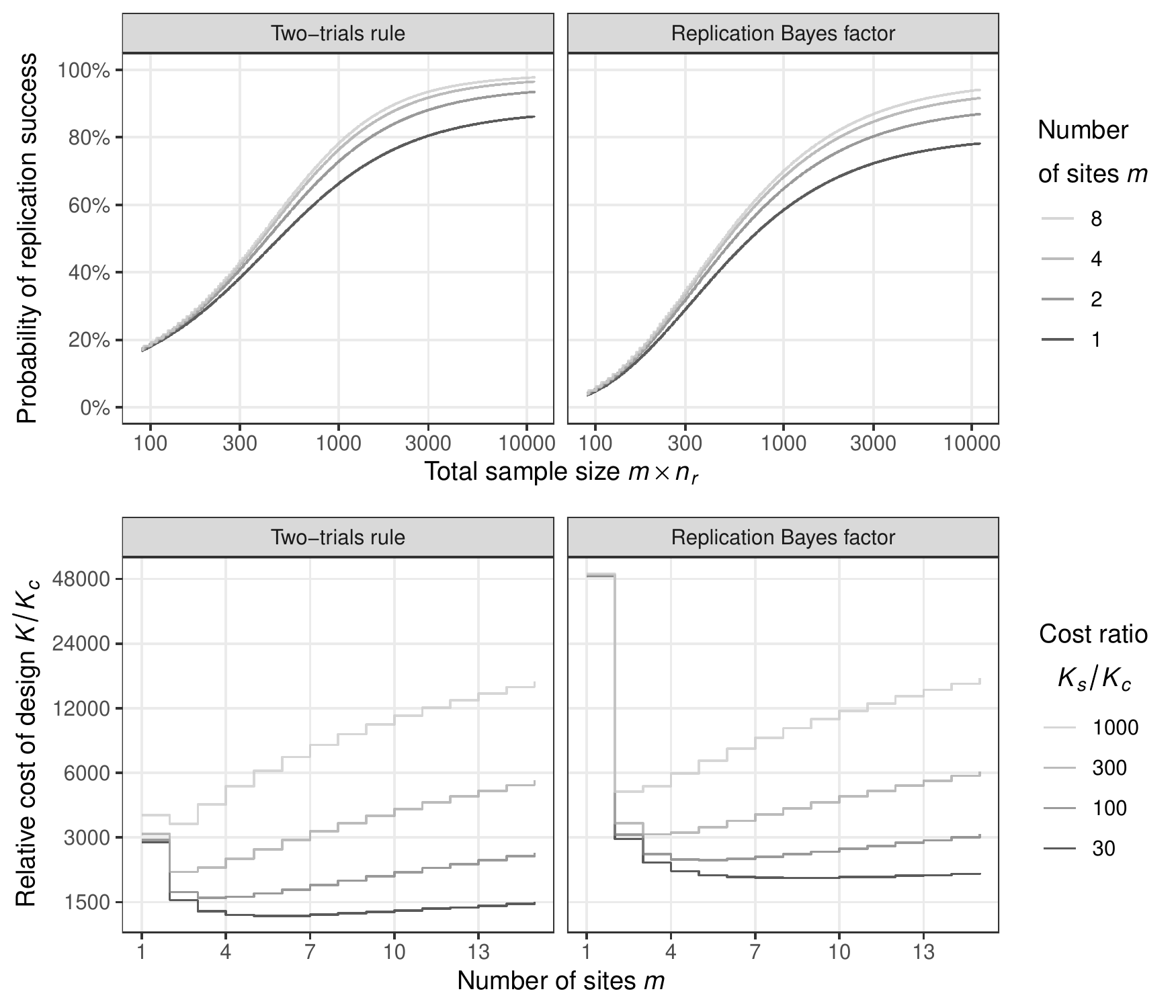} 
\end{knitrout}
\caption{The top plots show the probability of replication success based on the
  two-trials rule at level $\alpha = 0.025$ (left) and the
  replication Bayes factor at level $\gamma = 1/10$ (right) as a
  function of the total sample size and for different number of sites $m$ for
  data from the ``Labels'' experiment. A design prior with
  heterogeneity $\tau = 0.05$ and flat initial prior for the
  effect size $\theta$ is used. The same heterogeneity value is assumed in the
  analysis of the replications. 
  The bottom plot shows the total cost $K$ of the design (relative to the cost
  of a single sample $K_{c}$) as a function of the number of sites $m$ and for
  different site costs $K_{s}$. The sample size of each design corresponds to a
  target probability of replication success
  $1 - \beta = 80$\%.}
\label{fig:multisite}
\end{figure}

Figure~\ref{fig:multisite} illustrates multisite SSD for the
``Labels'' experiment from \citet{Protzko2020} for planned analyses
based on the two-trials rule and the replication Bayes factor (see the
supplement for details on the multisite extension of these two methods). As for
singlesite SSD, we use the design prior based on a flat initial prior for the
effect size and taking into account heterogeneity
($\tau = 0.05$). The top plots show the probability of
replication success as a function of the total sample size $m \times n_{r}$ for
different number of sites $m$. We see that for the same total sample size a
larger number of sites increases the probability of replication success. For
instance, a total sample size of roughly 3000 is required to achieve an 80\%
target probability with one site for the two-trials rule, whereas only
approximately half as many samples are required for two sites.

However, focusing only on the total sample size ignores the fact that the cost
of an additional site is usually larger than the cost of an additional sample.
The bottom plot shows the total cost $K$ of a design (relative to the cost of
one sample $K_{c}$) whose sample size is determined for a target probability of
replication success $1 - \beta = 80$\%. We see that if
the cost of an additional site $K_{s}$ is not much larger than the cost of an
additional sample $K_{c}$, e.g., $K_{s}/K_{c} = 30$, the optimal number of sites
is $m = 5$ for the two-trials rule and $m = 8$ for the replication Bayes factor.
If an additional site is more costly the optimal number of sites is lower, e.g.,
if the cost ratio is $K_{s}/K_{c} = 300$, the optimal number of sites is $m = 2$
for the two-trials rule and $m = 3$ for the replication Bayes factor. This is
similar to the actually used number of sites $m = 3$ (counting only
external-replications), respectively, $m = 4$ (counting also the
internal-replication) from \citet{Protzko2020}.

\section{Discussion}
\label{sec:discussion}

We showed how Bayesian approaches can be used to determine the sample size of
replication studies based on all the available information and the associated
uncertainty. A key strength of the approach is that it can be applied to any
type of replication analysis method, Bayesian or non-Bayesian, as long as there
is a well-defined success region for the replication effect estimate. Methods
for assessing replication success which have not yet been adapted to Bayesian
design approaches in the normal-normal hierarchical model (or not even proposed)
can thus benefit from our methodology. For instance, our methods could easily be
applied to the ``dual-criterion'' from \citet{Rosenkranz2021}, which defines
replication success via simultaneous statistical significance and practical
relevance of the effect estimates from the original and replication studies.

There are some limitations and possible extensions: we have developed the
methodology for ``direct'' replication studies \citep{Simons2014}, which attempt
to replicate the conditions of the original study as closely as possible.
However, SSD methodology is also needed for ``conceptual'' replication or
``generalization'' studies, which may have systematic deviations from the
original study. While the heterogeneity variance in the design prior allows SSD
to account for effect size heterogeneity to some extent, more research is needed
to investigate how to account for systematic study variation. For the same
reason, it is unclear how our Bayesian design approach can be applied to a
``causal'' replication framework \citep{Steiner2019,Wong2021}, where the focus
is on the ability of the original and replication studies to estimate the same
causal estimand, rather than on similar study procedures. In addition, as in
standard meta-analysis, we assumed that the variances of the effect estimates
are known, which can sometimes be inadequate \citep{Jackson2018}. Specifying
priors also for the variances could better reflect the available uncertainty but
would come at the cost of reduced interpretability and increased computational
complexity. We also did not consider designs in which the replication data are
analyzed sequentially. Ideas from Bayesian sequential designs
\citep{Schoenbrodt2017, Stefan2022} or from adaptive clinical trials
\citep{Bretz2009} could be adapted to the replication setting, as in
\citet{Micheloud2020}. A sequential analysis of the replication data could
possibly increase the efficiency of the replication. An additional point is that
we assumed that the original study has been completed when planning the
replication study. One could also consider a scenario where both the original
and the replication study are planned simultaneously and adopt a ``project''
perspective \citep{Maca2002, Held2021}. In this case, however, no information
from the original study is available and the design prior must be specified
entirely based on external knowledge. Finally, researchers have limited
resources and may not be able afford a large enough sample size to achieve their
desired probability of replication success. In this situation, a reverse-Bayes
approach \citep{Held2021b} could be used to determine the prior for the effect
size required to achieve the desired probability of replication success based on
the maximally affordable sample size. Researchers can then judge whether or not
such prior beliefs are scientifically sensible, and decide whether to conduct
the replication study with their limited resources.


\section{Appendix: The BayesRepDesign R package}
\label{app:package}
The R package BayesRepDesign can be installed from the Comprehensive R Archive
Network (CRAN) by running the following command from an R console

\begin{knitrout}\footnotesize
\definecolor{shadecolor}{rgb}{0.969, 0.969, 0.969}\color{fgcolor}\begin{kframe}
\begin{alltt}
\hlkwd{install.packages}\hlstd{(}\hlstr{"BayesRepDesign"}\hlstd{)}
\end{alltt}
\end{kframe}
\end{knitrout}

\begin{flushleft} Once the package is installed, it can be loaded with

\begin{knitrout}\footnotesize
\definecolor{shadecolor}{rgb}{0.969, 0.969, 0.969}\color{fgcolor}\begin{kframe}
\begin{alltt}
\hlkwd{library}\hlstd{(}\hlstr{"BayesRepDesign"}\hlstd{)}
\end{alltt}
\end{kframe}
\end{knitrout}

To see an overview of the functionality of the package, run

\begin{knitrout}\footnotesize
\definecolor{shadecolor}{rgb}{0.969, 0.969, 0.969}\color{fgcolor}\begin{kframe}
\begin{alltt}
\hlkwd{help}\hlstd{(}\hlkwc{package} \hlstd{=} \hlstr{"BayesRepDesign"}\hlstd{)}
\end{alltt}
\end{kframe}
\end{knitrout}

The first step in Bayesian design of a replication study is to create a design
prior for the effect size $\theta$. We use the original effect estimate
$\that_{o} = 0.205$ and standard error $\sigma_{o} = 0.051$ from the ``Labels''
experiment along with a flat initial prior for $\theta$ (the default) and a
heterogeneity standard deviation of $\tau = 0.05$ as inputs to the
\texttt{designPrior} function

\begin{knitrout}\footnotesize
\definecolor{shadecolor}{rgb}{0.969, 0.969, 0.969}\color{fgcolor}\begin{kframe}
\begin{alltt}
\hlstd{dp} \hlkwb{<-} \hlkwd{designPrior}\hlstd{(}\hlkwc{to} \hlstd{=} \hlnum{0.205}\hlstd{,} \hlkwc{so} \hlstd{=} \hlnum{0.051}\hlstd{,} \hlkwc{tau} \hlstd{=} \hlnum{0.05}\hlstd{)}
\end{alltt}
\end{kframe}
\end{knitrout}

The resulting design prior object can be visualized with

\begin{knitrout}\footnotesize
\definecolor{shadecolor}{rgb}{0.969, 0.969, 0.969}\color{fgcolor}\begin{kframe}
\begin{alltt}
\hlkwd{plot}\hlstd{(dp)}
\end{alltt}
\end{kframe}

{\centering \includegraphics[width=0.9\textwidth]{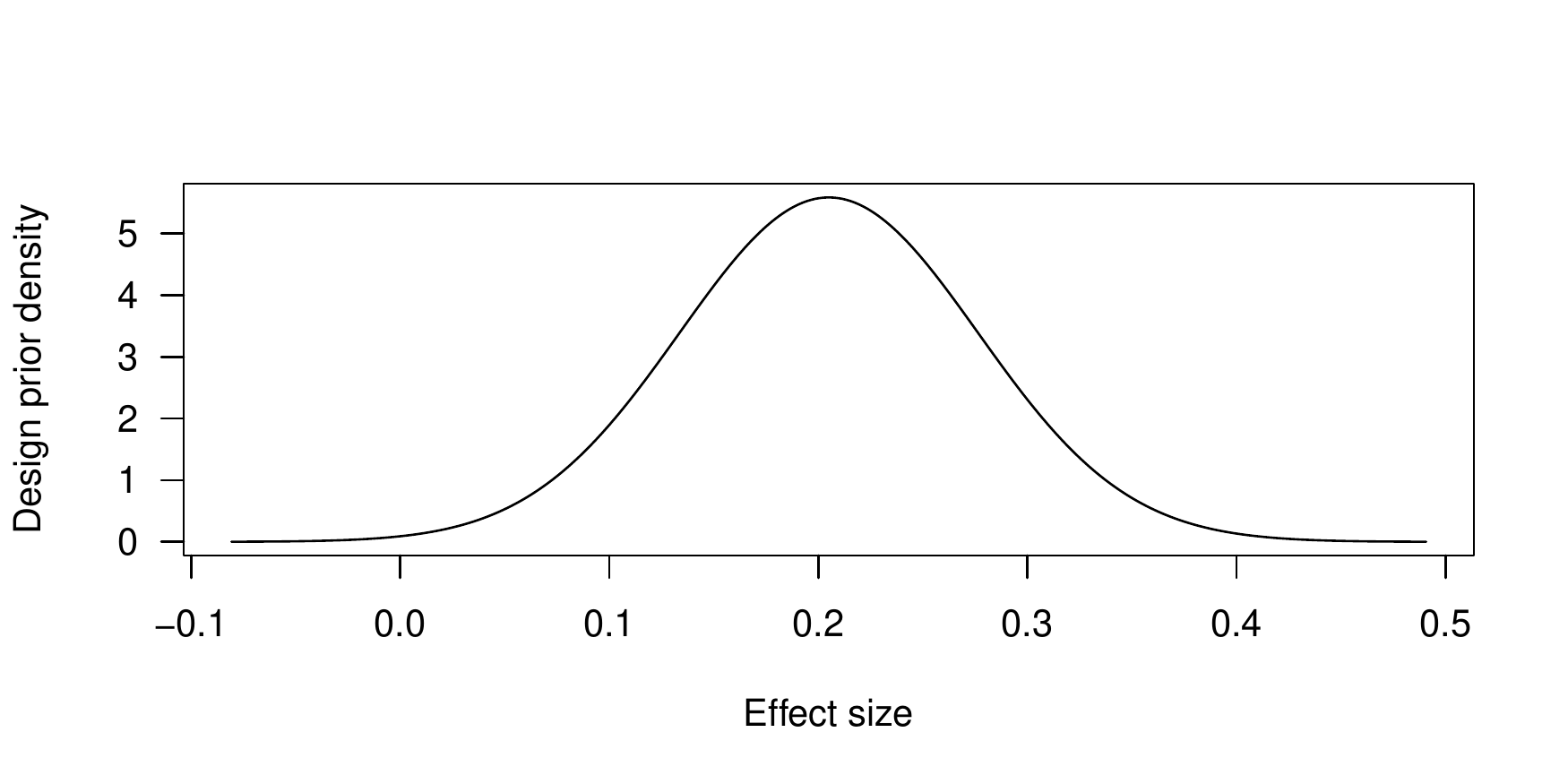} 

}

\end{knitrout}

The design prior can now be used to compute the probability of replication
success with the \texttt{pors} functions or to compute the replication standard
error with the \texttt{ssd} functions. Each analysis method discussed in this
paper has dedicated \texttt{pors} and \texttt{ssd} functions. For example,
\texttt{porsSig} can be used to compute the probability of replication success
defined by a significant replication $p$-value for a given replication standard
error, while \texttt{ssdSig} can be used to compute the replication standard
error required to achieve significance for a given target probability of
replication success. In the following, we will compute the replication standard
error for achieving replication success with a target probability of $80\%$.

\begin{knitrout}\footnotesize
\definecolor{shadecolor}{rgb}{0.969, 0.969, 0.969}\color{fgcolor}\begin{kframe}
\begin{alltt}
\hlstd{(ssd1} \hlkwb{<-} \hlkwd{ssdSig}\hlstd{(}\hlkwc{level} \hlstd{=} \hlnum{0.025}\hlstd{,} \hlkwc{dprior} \hlstd{= dp,} \hlkwc{power} \hlstd{=} \hlnum{0.8}\hlstd{))}
\end{alltt}
\begin{verbatim}
##        Bayesian sample size calculation for replication studies
##        ========================================================
## 
## success criterion and computation
## ------------------------------------------------------------------------
##   replication p-value <= 0.025 (exact computation) 
## 
## original data and initial prior for effect size
## ------------------------------------------------------------------------
##   to = 0.2 : original effect estimate
##   so = 0.051 : standard error of original effect estimate
##   tau = 0.05 : assumed heterogeneity standard deviation
##   N(mean = 0, sd = Inf) : initial normal prior
## 
## design prior for effect size
## ------------------------------------------------------------------------
##   N(mean = 0.2, sd = 0.071) : normal design prior
## 
## probability of replication success
## ------------------------------------------------------------------------
##   PoRS = 0.8 : specified
##   PoRS = 0.8 : recomputed with sr
## 
## required sample size
## ------------------------------------------------------------------------
##   sr = 0.059 : required standard error of replication effect estimate
##   c = so^2/sr^2 ~= nr/no = 0.74 : required relative variance / sample size
\end{verbatim}
\end{kframe}
\end{knitrout}

The output shows the relative variance $c = \sigma_{o}^{2}/\sigma_{r}^{2}$
which, assuming a standard error form $\sigma_{i} = \lambda/\surd{n_{i}}$, is
equal to the relative sample size $c = n_{r}/n_{o}$. The parameter $c$ thus
quantifies by how much the replication sample size $n_{r}$ must be
increased/decreased compared to the original sample size $n_{o}$. The
replication standard error can also be converted to an absolute sample size
using

\begin{knitrout}\footnotesize
\definecolor{shadecolor}{rgb}{0.969, 0.969, 0.969}\color{fgcolor}\begin{kframe}
\begin{alltt}
\hlkwd{se2n}\hlstd{(}\hlkwc{se} \hlstd{= ssd1}\hlopt{$}\hlstd{sr,} \hlkwc{unitSD} \hlstd{=} \hlnum{2}\hlstd{)}
\end{alltt}
\begin{verbatim}
## [1] 1137
\end{verbatim}
\end{kframe}
\end{knitrout}

This function assumes a unit standard deviation of $\lambda = 2$ for
the conversion which is a reasonable approximation of the unit standard
deviation for standardized mean differences and log odds/hazard/rate ratios for
balanced group designs \citep[Section 2.4]{Spiegelhalter2004}. However, more
exact conversions may be obtained by considering the exact form of the standard
error and solving for the sample size.
\end{flushleft}

The BayesRepDesign package can be easily extended to other replication analysis
methods than those for which dedicated functions are provided. To do so, users
need to define a function that returns the success region for the replication
effect estimate for a given replication standard error. The function is then
passed as an argument to the \texttt{ssd} function, which then numerically
determines the required standard error. The following code illustrates how the
significance method from earlier can be reimplemented in this way.

\begin{knitrout}\footnotesize
\definecolor{shadecolor}{rgb}{0.969, 0.969, 0.969}\color{fgcolor}\begin{kframe}
\begin{alltt}
\hlstd{sregionfunSig} \hlkwb{<-} \hlkwa{function}\hlstd{(}\hlkwc{sr}\hlstd{,} \hlkwc{alpha} \hlstd{=} \hlnum{0.025}\hlstd{) \{}
    \hlstd{za} \hlkwb{<-} \hlkwd{qnorm}\hlstd{(}\hlkwc{p} \hlstd{=} \hlnum{1} \hlopt{-} \hlstd{alpha)}
    \hlstd{sregion} \hlkwb{<-} \hlkwd{successRegion}\hlstd{(}\hlkwc{intervals} \hlstd{=} \hlkwd{cbind}\hlstd{(za}\hlopt{*}\hlstd{sr,} \hlnum{Inf}\hlstd{))}
    \hlkwd{return}\hlstd{(sregion)}
\hlstd{\}}
\hlstd{ssd2} \hlkwb{<-} \hlkwd{ssd}\hlstd{(}\hlkwc{sregionfun} \hlstd{= sregionfunSig,} \hlkwc{dprior} \hlstd{= dp,} \hlkwc{power} \hlstd{=} \hlnum{0.8}\hlstd{)}
\hlkwd{se2n}\hlstd{(}\hlkwc{se} \hlstd{= ssd2}\hlopt{$}\hlstd{sr,} \hlkwc{unitSD} \hlstd{=} \hlnum{2}\hlstd{)}
\end{alltt}
\begin{verbatim}
## [1] 1137
\end{verbatim}
\end{kframe}
\end{knitrout}

We see that this results in the same sample size as the \texttt{ssdSig}
function (which uses a closed-form solution).

\bibliography{bibliography}

\includepdf[pages=-]{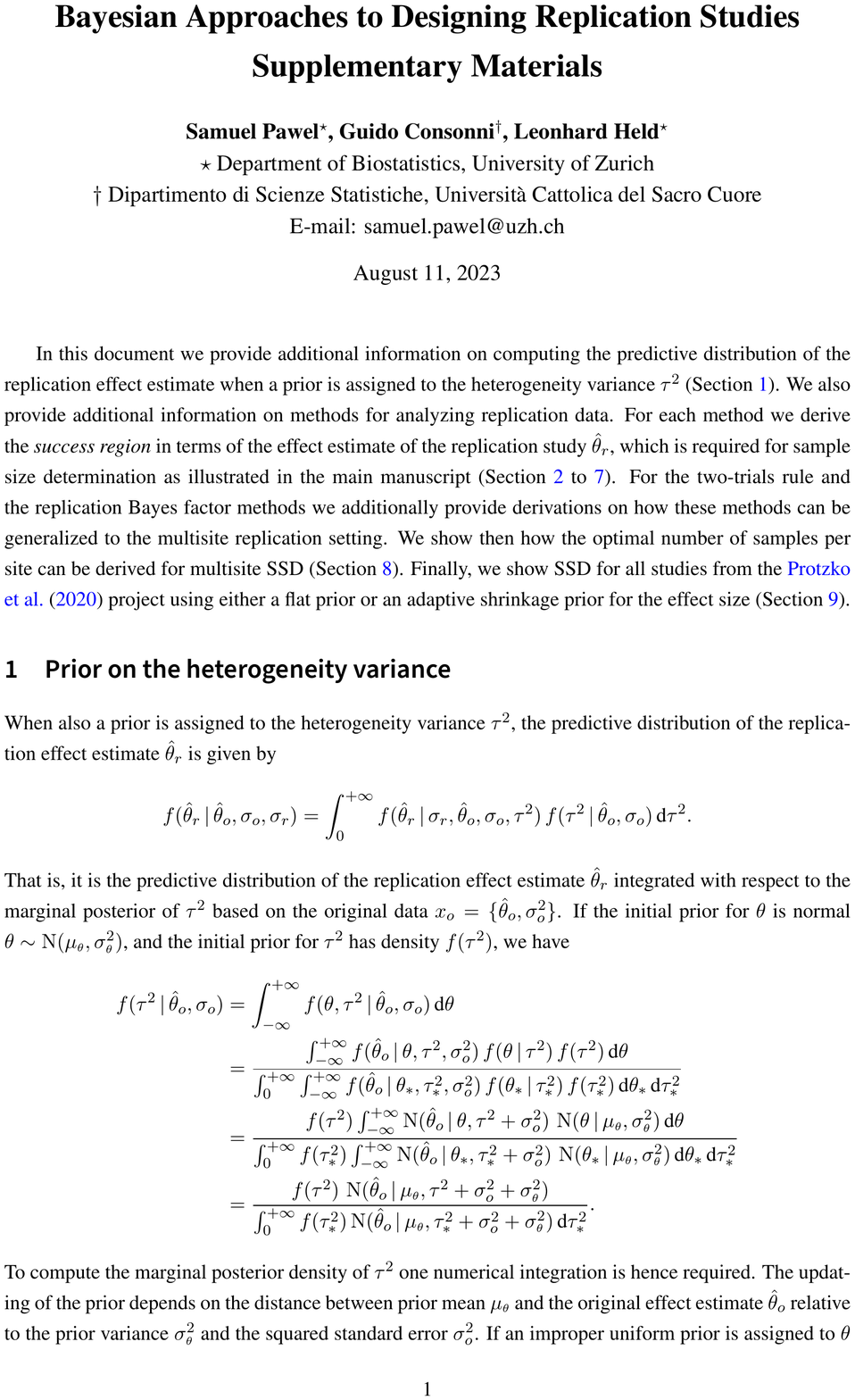}

\end{document}